\documentclass[11pt]{report}
\usepackage{natbib}
\usepackage{bibentry}
\usepackage{filecontents}
\usepackage[fleqn]{amsmath}
\usepackage{amssymb}
\usepackage[a4paper, total={6.5in, 9in}]{geometry}
\usepackage{fancyvrb}
\usepackage{spverbatim}
\usepackage{accents}
\usepackage{enumitem}
\usepackage{hyperref}
\usepackage{graphicx}
\usepackage{tabularx}

\hypersetup{ 
    colorlinks,
    citecolor=black,
    filecolor=black,
    linkcolor=black,
    urlcolor=black 
} 

\newcommand{\mb} {\mathbf}
\newcommand{\ms} {\boldsymbol}
\newcommand{\T}{^{\tiny \mathrm T}}
\newcommand{\TS}{^{{\mathrm T}/2}}
\newcommand{\ac}{\accentset}
\newcommand*\suptxt[1]{^{\textnormal{#1}}}
\newcommand*\subtxt[1]{_{\textnormal{#1}}}

\nobibliography*
\setlist[itemize]{itemsep=1pt, topsep=0pt}

\begin{document}

\title{EnKF-C user guide\\{\normalsize version 2.50.0}}

\author{Pavel Sakov}
\date{June 19, 2014 -- \today}

\maketitle
\thispagestyle{empty}

\clearpage

\tableofcontents

\clearpage

\chapter*{License}

EnKF-C

Copyright (C) 2014 Pavel Sakov and Bureau of Meteorology

Redistribution and use of material from the package EnKF-C, with or without
modification, are permitted provided that the following conditions are 
met:

   1. Redistributions of material must retain the above copyright notice, this
      list of conditions and the following disclaimer.
   2. The names of the authors may not be used to endorse or promote products
      derived from this software without specific prior written permission.

THIS SOFTWARE IS PROVIDED BY THE AUTHORS ``AS IS'' AND ANY EXPRESS OR IMPLIED 
WARRANTIES, INCLUDING, BUT NOT LIMITED TO, THE IMPLIED WARRANTIES OF
MERCHANTABILITY AND FITNESS FOR A PARTICULAR PURPOSE ARE DISCLAIMED. IN NO
EVENT SHALL THE AUTHORS BE LIABLE FOR ANY DIRECT, INDIRECT, INCIDENTAL, SPECIAL,
EXEMPLARY, OR CONSEQUENTIAL DAMAGES (INCLUDING, BUT NOT LIMITED TO, PROCUREMENT
OF SUBSTITUTE GOODS OR SERVICES; LOSS OF USE, DATA, OR PROFITS; OR BUSINESS
INTERRUPTION) HOWEVER CAUSED AND ON ANY THEORY OF LIABILITY, WHETHER IN
CONTRACT, STRICT LIABILITY, OR TORT (INCLUDING NEGLIGENCE OR OTHERWISE) ARISING
IN ANY WAY OUT OF THE USE OF THIS SOFTWARE, EVEN IF ADVISED OF THE POSSIBILITY
OF SUCH DAMAGE.

\chapter*{Introduction}
\addcontentsline{toc}{chapter}{Introduction}

EnKF-C aims to provide a compact generic framework for off-line data assimilation (DA) into large-scale layered geophysical models with the ensemble Kalman filter (EnKF).
Here ``compact'' has higher priority than ``generic''; that is, the code is not designed to cover every virtual possibility for the sake of it, but rather to be expandable in practical (from the author's point of view) situations.
Following are its other main features:
\begin{itemize}
\item coded in C for GNU/Linux platform;
\item model-agnostic;
\item can conduct DA either in EnKF, ensemble optimal interpolation (EnOI), or hybrid EnKF/EnOI modes;
\item permits multiple model grids;
\item can handle rectangular, curvilinear, or unstructured horizontal grids, z, sigma or hybrid vertical coordinates.
\end{itemize}

EnKF-C is available from \url{https://github.com/sakov/enkf-c}.
This user guide is a part of the EnKF-C package. 
It is also available from \url{http://arxiv.org/abs/1410.1233}.

The user guide has two main sections.
Section~\ref{ch:enkf} overviews the basics of the EnKF; Section~\ref{ch:enkf-c} provides technical description of EnKF-C.

\section*{Pre-requisites and limitations}

Following is the list of main prerequisites and limitations resulting from the design and algorithmic solutions adopted in EnKF-C:
\begin{itemize}
\item the model is assumed to be layered, so that the horizontal and vertical grids are independent of each other;
\item horizontal grids are assumed to be structured quadrilateral or unstructured;
\item the model output is assumed to be in NetCDF format, with $(x, y, z)$ dimension order (meaning $z$ is the ``slowest'', ``most outward'' variable);
\item the forecast observations are calculated off-line (outside the model) only;
\item there is no vertical localisation, so that one typically needs an ensemble of about 100 rather than 40 members.
\end{itemize}

\chapter{EnKF}
\label{ch:enkf}

\section{Kalman filter}

The Kalman filter (KF) is the underlying concept behind the EnKF.
It is rather simple if formulated as recursive least squares.

Consider the global (in time) nonlinear minimisation problem
\begin{align}
  \label{min-nonl}
  &\{\mb x_i^a\}_{i=1}^k = \arg \underset{\{\mb x_i\}_{i = 1}^k}\min J_k(\mb x_1, \dots, \mb x_k),\\
  &J_k(\mb x_1, \dots, \mb x_k)  = \|\mb x_1 - \mb x_1^f\|^2_{(\mb P_1^f)^{-1}}
  + \sum_{i = 1}^k \|\mb y_i - \mathcal H_i(\mb x_i)\|^2_{(\mb R_i)^{-1}}
  + \sum_{i = 2}^k \|\mb x_i - \mathcal M_i(\mb x_{i-1})\|^2_{(\mb Q_i)^{-1}}.
  \label{L-nonl}
\end{align}
Here $\{\mb x_i^a\}_{i=1}^k$ is a set of $k$ state vectors that minimise the cost function \eqref{L-nonl}; indices $i = 1,\dots,k$ correspond to a sequence of DA cycles, so that $\mb x_1$ is the estimated model state at the first cycle and $\mb x_k$ is the estimated model state at the last cycle; $\mb y_i$ are observation vectors; $\mathcal H_i$ are observation operators; $\mathcal M_i$ are model operators; $\mb P_1^f$ is the initial state error covariance; $\mb R_i$ are observation error covariances; $\mb Q_i$ are model error covariances; the norm notation $\|\mb x\|^2_{\mb B} \equiv \mb x\T \mb B \mb x$ is used; and $(\cdot)\T$ denotes matrix transposition.

The minimisation problem (\ref{min-nonl}, \ref{L-nonl}) is, generally, very complicated, but, luckily, has an exact solution in the \emph{linear} case; moreover, this solution is recursive.
Namely, assume that $\mathcal M$ and $\mathcal H$ are affine:
\begin{subequations}
  \label{lin}
  \begin{align}
    \label{lin-M}
    &\mathcal M_i(\mb x^{(1)}) - \mathcal M_i(\mb x^{(2)}) = \mb M_i \, (\mb x^{(1)} - \mb x^{(2)}),\\
    \label{lin-H}
    &\mathcal H_i(\mb x^{(1)}) - \mathcal H_i(\mb x^{(2)}) = \mb H_i \, (\mb x^{(1)} - \mb x^{(2)}),
  \end{align}
\end{subequations}
where $\mb x^{(1)}, \mb x^{(2)}$ are arbitrary model states, and $\mb M_i,\, \mb H_i = \mathrm{Const}$.
Then the cost function \eqref{L-nonl} becomes quadratic and can be written in canonical form in regard to $\mb x_k$:
\begin{align*}
   J_k(\mb x_1, \dots, \mb x_k) = \|\mb x_k - \mb x_k^a\|^2_{(\mb P_{k}^a)^{-1}} + \tilde{{J}}_{k-1}(\mb x_1,\dots,\mb x_{k-1}),
\end{align*}
so that
\begin{align}
  \label{min-xk}
  \underset {\{\mb x_i\}_{i = 1}^{k - 1}}{\min} J_k(\mb x_1, \dots, \mb x_k) = \|\mb x_k - \mb x_k^a\|^2_{(\mb P_{k}^a)^{-1}} + \mathrm{Const}.
\end{align}
(\emph{Proposition}) Then
\begin{align}
  \label{min-xkp1}
    \underset {\{\mb x_i\}_{i = 1}^{k}}{\min} J_{k+1}(\mb x_1, \dots, \mb x_k, \mb x_{k+1}) = \|\mb x_{k+1} - \mb x_{k+1}^a\|^2_{(\mb P_{k+1}^a)^{-1}} + \mathrm{Const},
\end{align}
where
\begin{subequations}
  \label{kf-an}
  \begin{align}
    \label{kf-an-x}
    & \mb x_{k+1}^a = \mb x_{k+1}^f + \mb K_{k+1} \left [\mb y_{k+1} - \mathcal H_{k+1}(\mb x_{k+1}^f) \right ],\\
    \label{kf-an-P}
    & \mb P_{k+1}^a = (\mb I - \mb K_{k+1} \mb H_{k+1}) \mb P_{k+1}^f,\\
    \nonumber
    & \hspace{-0.5cm} \text{where}\\
    \label{kf-K}
    & \mb K_{k+1} \equiv \mb P^f_{k+1} (\mb H_{k+1})\T \left [\mb H_{k+1} \mb P_{k+1}^f (\mb H_{k+1})\T + \mb R_{k+1} \right ]^{-1}
  \end{align}
\end{subequations}
\vspace{-8mm}
\begin{subequations}
  \label{kf-for}
  \begin{align}
    \label{prop-x}
    \nonumber
    & \hspace{-0.5cm} \text{and}\\
    & \mb x^f_{k+1} = \mathcal M_{k+1}(\mb x_k^a),\\
    \label{prop-P}
    & \mb P_{k+1}^f = \mb M_{k+1} \mb P_k^a (\mb M_{k+1})\T + \mb Q_{k+1}.
  \end{align}
\end{subequations}
This solution is known as the Kalman filter \citep[KF,][]{kal60}.
Equations \eqref{kf-for} describe advancing the system in time and represent the stage commonly called ``forecast'', while equations \eqref{kf-an} describe assimilation of observations and represent the stage called ``analysis''.
The superscripts $f$ and $a$ are used hereafter to refer to the forecast and analysis variables, respectively.
The forecast and analysis model state estimates $\mb x^f$ and $\mb x^a$ are commonly called (simply) forecast and analysis.
Matrix $\mb K$ is called Kalman gain.

\begin{figure}
  \centering
  \includegraphics[width = 0.7\textwidth]{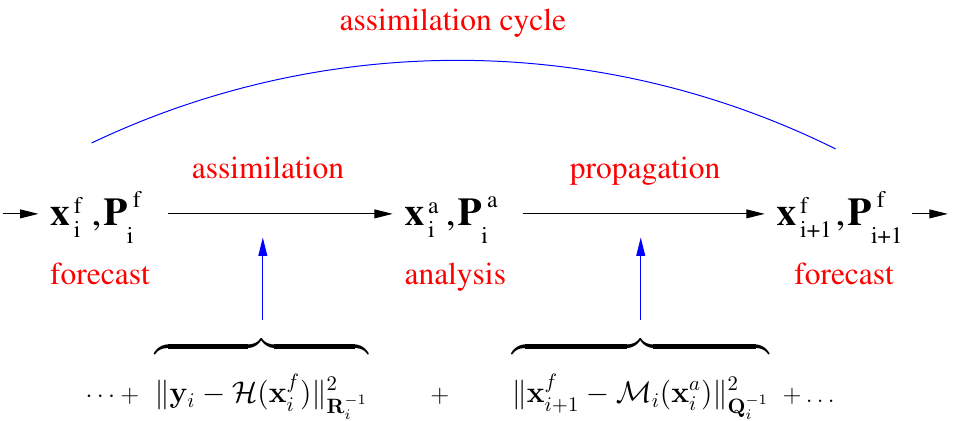}
  \caption{Data assimilation cycle of the Kalman filter.}
  \label{fig:dacycle}
\end{figure}

The recursive character of the KF makes it possible to consider solving the minimisation problem \eqref{L-nonl} as a sequence of forecasts and analyses, as shown in Fig.~\ref{fig:dacycle}.
Together an assimilation and the following propagation (or a propagation and the following assimilation) are referred to as assimilation cycle.

There are a few things to be noted about the KF:
\vspace{-3mm}
\begin{enumerate}
\item It follows from the Kalman filter (equations \ref{min-xk}-\ref{kf-for}) that the state of the DA system (SDAS) $\mb X$ is carried by the estimated model state vector and model state error covariance:
\begin{align}
  \label{sdas}
  \mb X_k = \{\mb x_k, \mb P_k\}.
\end{align}
\item The KF provides solution for the \emph{last} analysis, corresponding to $\mb x_k^a$ in ($\ref{min-nonl}$) (or, with a minor re-formulation, to the last forecast); finding the full (global in time) solution requires application of the Kalman \emph{smoother} (KS).
Both the KF and KS can be derived by decomposition of the positive (semi)definite quadratic function \eqref{L-nonl}.
\item Because the SDAS represents a (part of a) solution of the global least squares problem, it does not depend on the order in which observations are assimilated or on their grouping.
\item Ditto, the SDAS does not depend on a linear non-singular transform of the model state in the sense that the forward and inverse transforms commute with the evolution of the DA system.
\item Solution (\ref{kf-for}, \ref{kf-an}) can be \emph{used} in a nonlinear case by approximating
\begin{align*}
  &\mb M_{i} \leftarrow \nabla \mathcal M_i(\mb x_{i-1}^a),\\
  &\mb H_{i} \leftarrow \nabla \mathcal H_i(\mb x_i^f),
\end{align*}
in which case it is called the extended Kalman filter (EKF).
\end{enumerate}

\section{EnKF}

The standard form of the KF (\ref{kf-for}, \ref{kf-an}) is not necessarily the most convenient or suitable one in practice.
The corresponding algorithms can be prone to losing the positive definiteness of the state error covariance $\mb P$ due to rounding errors; and more importantly, explicit use of $\mb P$ makes these algorithms non-scalable in regard to the model state dimension.

Both these immediate problems can be addressed with the ensemble Kalman filter (EnKF).
In the EnKF the SDAS is carried by an ensemble of $m$ model states $\mb E$, which can be split into ensemble mean and ensemble anomalies:
\begin{align}
  \label{sdas-enkf}
  \mb X = \{\mb E\} = \{\mb x, \mb A\}.
\end{align}
It is related to the SDAS of the KF \eqref{sdas} as follows:
\begin{subequations}
  \label{E}
  \begin{align}
    \label{x}
    &\mb x = \frac{1}{m} \mb E \,\mb 1,\\
    \label{P}
    &\mb P = \frac{1}{m - 1} \mb A \mb A\T,\\
    \label{A}
    &\mb A \equiv \mb E - \mb x \mb 1\T,
  \end{align}
\end{subequations}
where $\mb 1$ is a vector with all elements equal to 1.
The above means that the model state estimate is given by the ensemble mean, while the model state error covariance $\mb P$ is implicitly represented by the ensemble anomalies $\mb A$ via the factorisation \eqref{P}.

Representing the state error covariance via ensemble anomalies yields a number of numerical benefits.
In large-scale geophysical systems the state size ($\sim 10^5-10^9$) makes it impossible to store and manipulate the state error covariance $\mb P$ directly.
At the same time it is often/typically possible to represent essential variability via an ensemble of much smaller size ($\sim 10^2$) and manipulate $\mb P$ implicitly via operations with $\mb A$.
Further, using $\mb A$ ensures positive semidefiniteness of $\mb P$.

Storing the SDAS via an ensemble of model states is the first essential feature of the EnKF.
In theory, one could use it in an implementation of the KF along with explicitly calculated Jacobians $\mb M$ and $\mb H$ in equations \eqref{kf-for} and \eqref{kf-an}.
The EnKF makes a further step and uses the ensemble form of the SDAS for a derivative-less formulation of the KF.
Moreover, it approximates derivatives using ensemble of finite spread that characterises the estimated uncertainty in the state:
\begin{subequations}
  \label{der-less-enkf}
  \begin{align}
    &\mb E \quad &\leftarrow& \quad \mb x \mb 1 \T  + \mb A \hspace{8cm}\\
    &\mathcal H (\mb x) \quad &\to& \quad \mathcal H (\mb E) \, \mb 1 / m\\
    &\mb H \mb A \quad &\to& \quad \mathcal H(\mb E)\left(\mb I - \mb 1 \mb 1\T / m\right)\\
    &\mathcal M (\mb x) \quad &\to& \quad \mathcal M (\mb E) \, \mb 1 / m\\
    &\mb M \mb A \quad &\to& \quad \mathcal M(\mb E)\left(\mb I - \mb 1 \mb 1\T / m\right)\\
    &\mb H \mb M \mb A &\to& \quad \mathcal H \circ \mathcal M(\mb E) \left(\mb I - \mb 1 \mb 1\T /m \right).
  \end{align}
\end{subequations}

This formulation is not the only one possible; one might also use the finite difference approximations:
\begin{subequations}
  \label{der-less-ekf}
  \begin{align}
    &\mb E \quad &\leftarrow& \quad \mb x \mb 1 \T  + \varepsilon \mb A \hspace{8cm}\\
    &\mb H \mb A \quad &\to& \quad \mathcal H(\mb E)\left(\mb I - \mb 1 \mb 1\T / m\right) / \varepsilon\\
    &\mb M \mb A \quad &\to& \quad \mathcal M(\mb E)\left(\mb I - \mb 1 \mb 1\T / m\right) / \varepsilon\\
    &\mb H \mb M \mb A \quad &\to& \quad \mathcal H \circ \mathcal M(\mb E)\left(\mb I - \mb 1 \mb 1\T / m\right) / \varepsilon.
  \end{align}
\end{subequations}
In this form the filter represents a derivative-less ensemble formulation of the EKF.
In practice the difference between the EnKF formulation \eqref{der-less-enkf} and the EKF formulation \eqref{der-less-ekf} is that the latter is more sensitive to small-scale variability and therefore more prone to instability, similar to the difference in behaviour of the Newton and secant methods.

The forecast stage of the EnKF involves just propagating each ensemble member:
\begin{align}
  \label{enkf-for}
  \mb E_i^f = \mathcal M_i(\mb E_{i-1}^a).
\end{align}
This is a remarkably simple equation compared to the KF forecast equations \eqref{kf-for}, even though the model error still needs to be accounted for in some way.
Because propagation of each ensemble member is independent from the other members, the forecast stage in the EnKF is naturally parallelisable.

One way to handle model error in the EnKF is to include stochastic model error into the model operator in \eqref{enkf-for}.
(This would make it different to the model operator in the KF, which is deterministic.)
Another option is to use the multiplicative inflation.
The third option is to mimic the treatment of model error in the KF, although this would require the ``rank reduction'' \citep[eq. 28]{ver97a} to prevent increasing the ensemble size.

At the analysis stage one has to update the ensemble mean and ensemble anomalies to match \eqref{kf-an}.
This involves handling ensemble as a whole, which is different to the forecast stage, when each ensemble member is propagated individually.

Note that factorisation \eqref{P} is not unique: if $\mb A$ satisfies \eqref{P}, then $\tilde {\mb A} = \mb A \mb U$, where $\mb U$ is an arbitrary orthonormal matrix $\mb U \mb U\T = \mb I$, also satisfies \eqref{P}.
However, $\tilde{\mb A}$ should not only factorise $\mb P$, but also remain an ensemble anomalies matrix, $\tilde{\mb A} \mb 1 = \mb 0$.
This requires an additional constraint $\mb U \mb 1 = \mb 1$.
Summarising, if $\mb E = \mb x \mb 1\T + \mb A$ is an ensemble that satisfies \eqref{E}, then ensemble
\begin{align}
  \label{redraw}
  \tilde{\mb E} = \mb x \mb 1\T + \mb A \mb U^p, \quad \mb U^p:\ \mb U^p (\mb U^p)\T = \mb I,\ \mb U^p \mb 1 = \mb 1
\end{align}
also satisfies \eqref{E}. 
If $\mb E$ is full rank (i.e. $\mathrm{rank}(\mb E) = \mathrm{min}(m, n)$, where $n$ is the state dimension), then each unique $\mb U^p$ generates a unique ensemble, and \eqref{redraw} describes all possible ensembles matching a given SDAS of the KF.
Such transformation of the ensemble is called ensemble \emph{redrawing}.
In the linear case (i.e. for affine model and observation operators) redrawing of the ensemble in the EnKF does not affect evolution of the underlying KF; and conversely, in the nonlinear case the redrawing does indeed affect evolution of the underlying KF.

\section{EnKF analysis}

In this section we will give a brief overview of solutions for EnKF analysis and describe the particular schemes used in EnKF-C.

\subsection{Overview}
\label{sec:enkf-overview}

In the ``baseline'' EnKF (full-rank ensemble, no localisation) the analysed SDAS matches that of the KF, although the algebraic side is indeed different.
The update of the ensemble mean is generally straightforward, in accordance with that in the KF \eqref{kf-an-x}.
The details may depend on the chosen algorithm to achieve better numerical efficiency (see Sec.~\ref{sec:numerical}).

The update of ensemble anomalies can be done either via a right-multiplied or left-multiplied (or post/pre-multiplied) transform of the ensemble anomalies:
\begin{align}
  \label{T-left}
  \mb A^a = \mb T_L \,\mb A^f,
\end{align}
or
\begin{align}
  \label{T-right}
  \mb A^a = \mb A^f \, \mb T_R.
\end{align}
$\mb T_L$ and $\mb T_R$ are referred to hereafter as left-multiplied and right-multiplied ensemble transform matrices (ETMs), respectively.
Note that to preserve the ensemble mean $\mb T_R$ has to satisfy $\mb T_R \mb 1 = \alpha \mb 1$, where $\alpha$ is an arbitrary constant.
It follows from \eqref{redraw} that if $\mb T_R$ is a particular solution for the right-multiplied ETM, then (for a full rank ensemble) any other solution can be written as
\begin{align}
  \label{TR-all}
  \tilde{\mb T}_R = \mb T_R \mb U^p, \quad \mb U^p:\ \mb U^p (\mb U^p)\T = \mb I,\ \mb U^p \mb 1 = \mb 1.
\end{align}
(More generally, one could write $\tilde{\mb T}_R = \mb T_R (\mb U^p + \mb 1 \ms a\T)$, where $\ms a$ is an arbitrary vector, but this additional term does not change the ensemble.)

Similarly, the analysis increment can be represented as a linear combination of the forecast ensemble anomalies:
\begin{align}
  \label{w}
  \mb x^a = \mb x^f + \mb A^f \mb w.
\end{align}

Equations \eqref{T-right} and \eqref{w} can be combined into a single transform of the ensemble:
\begin{align}
  \label{x5-def}
  & \mb E^a = \mb E^f \mb X_5,\\
  \label{x5-raw}
  & \mb X_5 = \frac{1}{m} \mb 1 \mb 1\T + \left( \mb I - \frac{1}{m} \mb 1 \mb 1\T\right) \left( \mb w \mb 1\T + \mb T_R\right)
  = \frac{1}{m} \mb 1 \mb 1\T + \mb w \mb 1\T + \left( \mb I - \frac{1}{m} \mb 1 \mb 1\T \right) \mb T_R,
\end{align}
as $\mb 1\T \mb w = 0$. 
When $\mb T_R:\ \mb T_R\T = \mb T_R,\ \mb T_R \mb 1 = \mb 1$ (which is the case for e.g. the ETKF and DEnKF), $\mb 1\T \mb T_R = \mb 1\T$, and \eqref{x5-raw} simplifies to
\begin{align}
  \label{x5}
  & \mb X_5 = \mb w \mb 1\T + \mb T_R.
\end{align}
The designation $\mb X_5$ is used for historic reasons, following \citet{eve03a}.

\subsection{Some schemes}

As follows from the previous section, there are multiple solutions for the ETM that match the KF covariance update equation \eqref{kf-an-P}; however the particular solutions may have different properties in practice due to the DAS nonlinearity, their algorithmic convenience, or their robustness in suboptimal conditions.
This section provides some background for the schemes used in EnKF-C:
\begin{itemize}
\item ETKF;
\item DEnKF.
\end{itemize}

\subsubsection{ETKF}

It is easy to show using the definition of $\mb K$ \eqref{kf-K} and matrix shift lemma \eqref{shift} that 
\begin{align*}
    (\mb I - \mb K \mb H) \, \mb P^f = (\mb I - \mb K \mb H)^{1/2} \; \mb P^f \; (\mb I - \mb K \mb H)\TS,
\end{align*}
which yields the following solution for the left-multiplied ETM:
\begin{align}
  \label{T-left1}
  \mb T_L = (\mb I - \mb K \mb H)^{1/2}
\end{align}
\citep{sak08b}, that is
\begin{align}
  \tag{\ref{T-left1}a}
  \label{T-left1a}
  \mb A^a =  (\mb I - \mb K \mb H)^{1/2} \mb A^f.
\end{align}
Hereafter by $\mb X^{1/2}$ we denote the unique positive definite square root of a positive definite (generally, non-symmetric) matrix $\mb X$, defined as $\mb X^{1/2} = \mb V \mb L^{1/2} \mb V^{-1}$, where $\mb X = \mb V \mb L \mb V^{-1}$ is the eigenvalue decomposition of $\mb X$.
By ``matrix shift lemma'' we refer to the following identity:
\begin{align}
  \label{shift}
  \mathcal F(\mb A \mb B) \, \mb A = \mb A \, \mathcal F(\mb B \mb A),
\end{align}
where $\mathcal F$ is an arbitrary function expandable into Taylor series.
Rewriting \eqref{T-left1a} as
\begin{align*}
  \mb A^a = \left[ \mb I - \frac{1}{m-1} \mb A^f (\mb H \mb A^f)\T  (\mb H \mb P^f \mb H\T + \mb R)^{-1} \mb H \right]^{1/2} \mb A^f
\end{align*}
and using the matrix shift lemma, we obtain:
\begin{align*}
  \mb A^a = \mb A^f \left[ \mb I - \frac{1}{m-1} (\mb H \mb A^f)\T  (\mb H \mb P^f \mb H\T + \mb R)^{-1} \mb H \mb A^f \right]^{1/2}
\end{align*}
which yields the corresponding to \eqref{T-left1} right-multiplied ETM:
\begin{align}
  \label{T-right1}
  \mb T_R = \left[ \mb I -  \frac{1}{m-1} (\mb H \mb A^f)\T  (\mb H \mb P^f \mb H\T + \mb R)^{-1} \mb H \mb A^f \right]^{1/2}
\end{align}
\citep{eve04a}.
Applying the matrix inversion lemma
\begin{align}
  \label{inv}
  (\mb A + \mb U \mb L \mb V)^{-1} = \mb A^{-1} - \mb A^{-1} \mb U (\mb L^{-1} + \mb V \mb A^{-1} \mb U)^{-1} \mb V \mb A^{-1},
\end{align}
\eqref{T-left1} can be transformed to:
\begin{align}
  \label{T-left2}
  \mb T_L = (\mb I + \mb P^f \mb H\T \mb R^{-1} \mb H)^{-1/2}
\end{align}
\citep{sak11a}; and applying the matrix shift lemma yields the corresponding right-multiplied ETM:
\begin{align}
  \label{etkf}
  \mb T_R = \left[\mb I +  \frac{1}{m-1} (\mb H \mb A^f)\T \mb R^{-1} \mb H \mb A^f \right]^{-1/2},
\end{align}
commonly known as the ensemble transform Kalman filter, or ETKF \citep{bis01a}.

{
  \scriptsize
  {\bf Historic reference.} Another (and probably the first) solution for $\mb T_R$ equivalent to \eqref{T-right1} and \eqref{etkf} was found by \citet{and68a}:
  \setlength{\abovedisplayskip}{1pt}
  \setlength{\belowdisplayskip}{3pt}
  \begin{align}
    \label{andrews}
    \mb T_R = \mb I - \frac{1}{m-1} (\mb H \mb A^f)\T \mb M^{-1/2} \left(\mb M^{1/2} + \mb R^{1/2}\right)^{-1} \mb H \mb A^f,
  \end{align}
  where $\mb M \equiv \mb H \mb P^f \mb H\T + \mb R$.
}

{
  \scriptsize
  {\bf Correction.} The Andrews' solution turns out to be valid only for diagonal $\mb R$.
  The correct solution of form \eqref{andrews} can be obtained from \eqref{etkf} by using identity $\mb M^{-1/2} = \mb I - (\mb M + \mb M^{1/2})^{-1}(\mb M - \mb I)$:
  \setlength{\abovedisplayskip}{1pt}
  \setlength{\belowdisplayskip}{3pt}
  \begin{subequations}
    \label{bocquet}
    \begin{align}
      \mb T_R &= \mb I - \left(\mb M_m + \mb M_m^{1/2}\right)^{-1} \mb S\T \mb S\\
      \label{bocquet-p}
      &= \mb I - \mb S\T \left(\mb M_p + \mb M_p^{1/2}\right)^{-1} \mb S,
    \end{align}
 \end{subequations}
  where $\mb M_m \equiv \mb I + \mb S\T \mb S$, $\mb M_p \equiv \mb I + \mb S \mb S\T$, $\mb S \equiv \mb R^{-1/2} \mb H \mb A^f / (m - 1)^{1/2}$ \citep{boc16a}.

  \eqref{bocquet-p} can be written as
  \begin{align}
    \label{andrews-correct}
    \mb T_R = \mb I - \frac{1}{m-1} (\mb H \mb A^f)\T \hat{\mb M}^{-1/2} \left(\hat{\mb M}^{1/2} + \mb R^{1/2}\right)^{-1} \mb H \mb A^f,
  \end{align}
  where $\hat{\mb M}^{1/2} \equiv \mb R^{1/2} (\mb R^{-1/2} \mb M \mb R^{-1/2})^{1/2}$.
\eqref{andrews-correct} coincides with the Andrews' solution \eqref{andrews} if $\hat{\mb M}^{1/2} = \mb M^{1/2}$, which is the case for diagonal $\mb R$.
}

Equations (\ref{T-right1}, \ref{etkf}, \ref{bocquet}) yield algebraically different expressions for the (unique) symmetric right-multiplied solution.
Apart from being the only symmetric solution, it also represents the minimum distance solution for the ensemble anomalies: its ensemble of analysed anomalies is closer to the ensemble of forecast anomalies with the inverse forecast (or analysis) covariance as the metric than any other ensemble of analysed anomalies given by \eqref{TR-all} \citep{ott03a}.
This means that in the above sense the symmetric right-multiplied solution preserves the identities of ensemble members during analysis in the best possible way.

Note that while the left-multiplied solutions (\ref{T-left1}, \ref{T-left2}) correspond to the symmetric right-multiplied solution, they are not symmetric.

In a typical DAS with a large scale model one can expect $m = 100$, $p = 10^3-10^7$, $n = 10^6-10^9$; that is
\begin{align}
  m \ll p \ll n.
\end{align}
Therefore, considering the size of ETMs ($n \times n$ for left-multiplied ETMs and $m \times m$ for right-multiplied ETMs), only right-multiplied solutions are suitable for use with large scale models.
The ETKF solution \eqref{etkf} represents the most popular option due to its simple form and numerical effectiveness: for a diagonal $\mb R$, it only requires to calculate inverse square root of a symmetric $m \times m$ matrix.
Also, along with the left-multiplied solution \eqref{T-left2}, it generally has better numerical properties than solutions \eqref{T-left1} and \eqref{T-right1} due to the fact that the inverse square root in it is calculated from the sum of a positive definite and a positive semi-definite matrices.

\subsubsection{DEnKF}

Assuming that $\mb K \mb H$ is small in some sense, one can approximate solution \eqref{T-left1} by expanding it into Taylor series about $\mb I$ and keeping the first two terms of the expansion:
\begin{align}
  \label{denkf}
  \mb T_L = \mb I - \frac{1}{2} \mb K \mb H.
\end{align}
This approximation is known as the deterministic ensemble Kalman filter, or DEnKF \citep{sak08a}.
It has a simple interpretation of using half of the Kalman gain for updating the ensemble anomalies; but apart from that the DEnKF often represents a good practical choice due to its algorithmic convenience and good performance in suboptimal situations.
The DEnKF is the default scheme in EnKF-C.

\subsection{Some numerical considerations}
\label{sec:numerical}

Instead of using the forecast ensemble observation anomalies $\mb H \mb A^f$ and innovation $\mb y - \mathcal H(\mb x^f)$ it is convenient to use their standardised versions:
\begin{align}
  \label{s}
  &\mb s = \mb R^{-1/2} \left[ \mb y - \mathcal H(\mb x^f) \right] / \sqrt{m - 1},\\
  \label{S}
  &\mb S = \mb R^{-1/2} \mb H \mb A^f / \sqrt{m - 1}.
\end{align}
Then
\begin{align}
  \label{w-Gs}
  \mb w = \mb G \mb s;
\end{align}
for the ETKF
\begin{align}
  \label{TR-ETKF}
  \mb T_R = (\mb I + \mb S\T \mb S)^{-1/2},
\end{align}
and for the DEnKF
\begin{align}
  \label{TR-DEnKF}
  \mb T_R = \mb I - \frac{1}{2} \mb G \mb S,
\end{align}
where
\begin{align}
  \label{G-m}
  \mb G &\equiv (\mb I + \mb S\T \mb S)^{-1} \mb S\T,\\
  \label{G-p}
  & = \mb S\T ( \mb I + \mb S \mb S\T)^{-1}.
\end{align}
Here \eqref{G-m} involves inversion of an $m \times m$ matrix, while \eqref{G-p} involves inversion of a $p \times p$ matrix.
Therefore, in the DEnKF it is possible to calculate $\mb w$ and $\mb T$ using a single inversion of either a $p \times p$ or $m \times m$ matrix, depending on the relation between the number of observations and the ensemble size.
The ETKF can exploit the smaller dimension in the same way by using form \eqref{bocquet} of \eqref{TR-ETKF}: EnKF-C calculates $\mb G$ and $\mb T_R = \mb I - (\mb M_m + \mb M_m^{1/2})^{-1} \mb S\T \mb S$ (or its $p \times p$ equivalent \eqref{bocquet-p}) from a single singular value decomposition (SVD) of either the $m \times m$ or the $p \times p$ matrix, whichever is smaller.
Therefore both schemes benefit when $p < m$; the DEnKF is somewhat more numerically effective because it requires only matrix inversion, which can be done via Cholesky decomposition instead of SVD.

\subsection{Ensemble reduction}
\label{sec:reduction}

A number of EnKF extensions, such as the EnKF/EnOI hybrid (Sec.~\ref{sec:hybrid-theory}), use a larger ensemble for analysis than for forecasting.
These extensions require an ensemble reduction step following the analysis.

Assume an initial forecast ensemble $\mb E^f$ of size $m$ that can be partitioned into the ensemble average $\mb x^f$ and ensemble anomalies $\mb A^f$:
\begin{align*}
  \mb E^f = \mb x^f \mb 1\T + \mb A^f.
\end{align*}
Further assume that these initial forecast anomalies are expanded into a larger ensemble of forecast anomalies $\tilde{\mb A}^f$ of size $\tilde{m} > m$:
\begin{align*}
  \mb A^f \to \tilde{\mb A}^f,
\end{align*}
which is subsequently used for the analysis.
The analysis yields an expanded ensemble of anomalies $\tilde{\mb A}^a$, which must then be reduced to the initial size $m$:
\begin{align*}
  \tilde{\mb A}^a \to \mb A^a.
\end{align*}
A natural way to conduct this ensemble reduction can be formulated via the left-multiplied ensemble transform:
\begin{align}
  \label{reduction_left}
  \mb A^a = \tilde{\mb T}_L \mb A^f,
\end{align}
where $\tilde{\mb T}_L$ is computed from the expanded forecast ensemble anomalies: $\tilde{\mb T}_L = \mb T_L(\tilde{\mb A}^f)$.

To be computationally feasible in large-scale systems, this approach must be reformulated for an ensemble update utilizing a right-multiplied transform.
This reformulation is straightforward if $\mb A^f$ is structurally embedded within $\tilde{\mb A}^f$.

If the initial ensemble $\mb A^f$ forms a submatrix of $\tilde{\mb A}^f$ such that $\tilde{\mb A}^f = [\mb A^f, \dots]$, the reduction in \eqref{reduction_left} is equivalent to updating the expanded ensemble and then retaining its first $m$ members as the final updated ensemble.
The equivalent operation for the right-multiplied transform update consists of truncating the transform matrix to its first $m$ columns: $\mb A^a = \tilde{\mb A}^f \tilde{\mb T}_R(:, 1:m)$.

A simple generalization of this case is when a scalar multiple of $\mb A^f$ forms a submatrix of $\tilde{\mb A}^f$ such that
\begin{align}
  \tilde{\mb A}^f = [k \mb A^f, \dots].
\end{align}
In this case, \eqref{reduction_left} is equivalent to updating the expanded ensemble, retaining the first $m$ members of the expanded updated ensemble, and applying inverse scaling:
\begin{align*}
  \tilde{\mb T}_L \mb A^f = k^{-1} \tilde{\mb T}_L \tilde{\mb A}^f(:, 1 : m).
\end{align*}
The equivalent operation for the right-multiplied update involves truncating the transform matrix along with the scaling:
\begin{align}
  \label{reduction_right}
  \mb A^a = k^{-1} \tilde{\mb A}^f \tilde{\mb T}_R(:, 1:m).
\end{align}

\section{Asynchronous DA}

Observations assimilated at each cycle in the KF are assumed to be made simultaneously at the time of assimilation.
In such cases, observations and DA method are referred to as \emph{synchronous}.
In reality, observations assimilated at a given cycle are made over some period of time called ``data assimilation window'' (DAW).
If the DA method accounts for the time of observations, observations and DA method are referred to as \emph{asynchronous}.

The EnKF can be naturally extended for asynchronous DA.
Let us consider the minimisation problem (\ref{min-nonl}, \ref{L-nonl}) in the case of \emph{perfect model} $\mb Q = 0$.
It becomes
\begin{align}
  \label{min-perf}
  &{\mb x_1^a} = \arg \min \; J(\mb x_1),\\
  \label{cost-perf}
  &J(\mb x_1)  = \|\mb x_1 - \mb x_1^f\|^2_{(\mb P_1^f)^{-1}} + \sum_{i = 1}^k \|\mb y_i - \mathcal H_i(\mb x_i)\|^2_{(\mb R_i)^{-1}},\\
  \label{model-perf}
  &\mb x_{i+1} = \mathcal M(\mb x_i),\quad i = 1,\dots,k-1.
\end{align}
Compared to the original problem, the dimensionality of the solution is much reduced due to relations \eqref{model-perf}, which mean that the model state at any time can be found by propagating the initial state: $\mb x_2 = \mathcal M_2(\mb x_1),\ \mb x_3 = \mathcal M_3 \circ \mathcal M_2(\mb x_1), \dots$.
The cost function \eqref{cost-perf} can then be written as
\begin{align}
  \label{cost-async}
  J(\mb x_1)  = \|\mb x_1 - \mb x_1^f\|^2_{(\mb P^f)^{-1}} + \|\mb y - \mathcal H \circ \mathcal M(\mb x_1)\|^2_{\mb R^{-1}},
\end{align}
where observations $\mb y$ represent the augmented observation vector: $\mb y = [\mb y_1\T,\dots,\mb y_k\T]\T$, $\mb R$ is the corresponding observation error covariance, and forward operator $\mathcal H \circ \mathcal M (\mb x_1)$ relates the initial state $\mb x_1$ to observations.

Note that usually only $\mathcal H$ is a function that depends on assimilated observations; now by introducing $\mathcal H \circ \mathcal M (\mb x) = \mathcal H[\mathcal M(\mb x)]$ we have to assume that $\mathcal M$ also depends on observations, propagating the initial state to the time of each observation.
It is also possible to interpret $\mathcal M(\mb x$) as the \emph{trajectory} starting from $\mb x$, while $\mathcal H$ maps it to observations.

Apart from the operator $\mathcal H \circ \mathcal M$, the cost function \eqref{cost-async} has the same form as that for a single DA cycle with synchronous observations.
Consequently, \emph{in the linear case} \eqref{lin} one can use solutions for $\mb w$ and $\mb T$ from Section~\ref{sec:numerical}, subject to extending definitions of $\mb s$ \eqref{s} and $\mb S$ \eqref{S} as follows:
\begin{align}
  \label{s-async}
  &\mb s = \mb R^{-1/2} \left[ \mb y - \mathcal H \circ \mathcal M (\mb x_1^f) \right] / \sqrt{m - 1},\\
  \label{S-async}
  &\mb S = \mb R^{-1/2} \, \mb H \circ \mb M \, \mb A_1^f / \sqrt{m - 1},
\end{align}
where $\mb H \circ \mb M$ is the tangent linear operator of $\mathcal H \circ \mathcal M$ about $\mb x_1^f$.
This means that to account for the time of observations in the EnKF one simply needs calculate innovation and forecast ensemble observation anomalies using ensemble at the time of each observation.
There are no specific restrictions on $\mb R$, so that in theory observation errors can be correlated in time.

The minimisation problem \eqref{min-perf},\eqref{cost-async} implies assimilation time $t = t_1$; however, in the linear case the standardised innovation and ensemble observation anomalies in form (\ref{s-async},\ref{S-async}) represent objects invariant to assimilation time: the reference to $\mb x_1$ is only needed to define forward operator $\mathcal H \circ \mathcal M$.
Consequently, the ensemble transform $\mb X_5$ calculated from $\mb s$ and $\mb S$ can be applied to ensemble at any particular time to yield (the same) analysed trajectories for the ensemble members: $\mathcal M (\mb E) \, \mb X_5 = \mathcal M (\mb E \, \mb X_5)$.
This time invariance of ensemble transforms can be used to update ensemble back in time using observations from future cycles without the need in backward model (\citealt[][Sec.~6]{eve00a}, \citealt[][app.~D]{eve03a}).

{
  \setlength{\abovedisplayskip}{2pt}
  \setlength{\belowdisplayskip}{2pt}
  \scriptsize
  {\bf Note.} The background term $\|\mb x_1 - \mb x_1^f\|^2_{(\mb P^f)^{-1}}$ in \eqref{cost-async} can be seen as accumulating the previous history of the system rather than characterising the initial uncertainty in the global problem.
  In this case it is natural to anchor it to the previous analysis:
  \begin{align}
    J(\mb x) = \|\mb x - \mb x^f\|^2_{(\mb P^f)^{-1}} + \|\mb y - \mathcal H \circ \mathcal M(\mb x)\|^2_{\mb R^{-1}}.
  \end{align}
  Here $\mb x^f$ is the forecast state at the start of the cycle obtained from the previous analysis, and $\mb P^f$ is the corresponding state error covariance.
  Minimising $J$ yields the analysed initial model state $\mb x^a$, which in turn yields the analysed trajectory.
  The analysed state error covariance is defined so that the analysed background term absorbs the observation term:
  \begin{align*}
    &\mb x^a, \mb P^a: \ J(\mb x) = \|\mb x - \mb x^a\|^2_{(\mb P^a)^{-1}} + \mathrm{Const}.
  \end{align*}
  This framework is a natural extension of the problem \eqref{min-nonl} in the linear, perfect-model case to continuous time; and is convenient for iterative minimisation.\par
}

\section{EnOI}

The EnOI, or ensemble optimal interpolation \citep{eve03a}, can be defined as the EnKF with static or, more generally, predefined, ensemble anomalies.
It can be summarised as follows:
\begin{align}
  \label{enoi-for}
  & \mb x^b_i = \mathcal M_i(\mb x^a_{i-1}),\\
  & \mb x^a_i = \mb x^b_i + \mb A^b \, \mb w_i,
\end{align}
where $\mb x^b$ is the forecast model state estimate referred to as background, and $\mb A^b$ is an ensemble of static, or background, anomalies; the corresponding state error covariance $\mb P^b$ is also often referred to as the background covariance.

The main incentive for using the EnOI is its low computational cost due to the integration of only one instance of the model.
Despite of the similarity with the EnKF, the EnOI is a rather different concept, as there is no global in time cost function associated with it.
Conceptually the EnOI is closer to 3D-Var, as both methods use static (anisotropic, multivariate) covariance.
It is an improvement on the optimal interpolation, which typically uses isotropic, homogeneous and univariate covariance.

In contrast to the EnKF, due to the use of a static ensemble the EnOI avoids potential problems related to the ensemble spread; but at the same time it does critically depend on the ensemble, while the EnKF with a stochastic model typically ``forgets'' the initial ensemble over time.

The EnOI can account for the time of observations by calculating innovation using forecast at observation time, as in \eqref{s-async}.
This approach is commonly known as ``first guess at appropriate time'', or FGAT.

\section{EnKF/EnOI hybrid}
\label{sec:hybrid-theory}

By an EnKF/EnOI hybrid, we understand a formulation in which forecast state error covariance is equal to the sum of a ``dynamic'' covariance carried by the EnKF ensemble, and a ``static'' covariance carried by one or more predefined ensembles of anomalies:
\begin{align}
  \label{hybrid_covariance}
  \mb P^f = \mb P\suptxt{dyn} + \sum_{s = 1}^N \gamma_s \mb P\suptxt{stat}_s,
\end{align}
where
\begin{align*}
  & \mb P\suptxt{dyn} = \frac{1}{m\subtxt{dyn} - 1}\mb A\suptxt{dyn} (\mb A\suptxt{dyn})\T,\\
  & \mb P\suptxt{stat}_s = \frac{1}{m_s - 1}\mb A\suptxt{stat}_s (\mb A\suptxt{stat}_s)\T, \quad s = 1, \dots, N,
\end{align*}
where $\mb A\suptxt{dyn}$ is the ensemble of dynamic anomalies of size $m\subtxt{dyn}$, $\mb A\suptxt{stat}_s$ is the $s$-th ensemble of static anomalies of size $m_s$, $N \ge 1$ is the number of static ensembles, and $\gamma_s > 0$ is the scaling coefficient of the $s$-th static ensemble.
The static anomalies are obtained by subtracting the ensemble mean from each static ensemble.
The added static covariance can be assumed to represent the model error covariance (matrices $\mb Q_i$ in \eqref{L-nonl} and $\mb Q_{k+1}$ in \eqref{prop-P}).

The forecast covariance $\mb P^f$ is then carried by the combined ensemble $\mb A^f$:
\begin{align}
  \label{hybrid_combined}
  & \mb A^f = \left[ k\subtxt{dyn} \mb A\suptxt{dyn}, k_1 \mb A\suptxt{stat}_1, \dots, k_N \mb A\suptxt{stat}_N \right],\\
  \nonumber
  & k\subtxt{dyn} = \left(\frac{m - N - 1}{m\subtxt{dyn} - 1}\right)^{1/2}, \quad k_s = \left(\gamma_s \frac{m - N - 1}{m_s - 1}\right)^{1/2},
\end{align}
where $m = m\subtxt{dyn} + \sum_{s=1}^N m_s$, so that
\begin{align}
  \label{cov_combined}
  \mb P^f = \frac{1}{m - N - 1} \mb A^f (\mb A^f)\T.
\end{align}

We use factor $m - N - 1$ instead of the usual $m - 1$ in \eqref{cov_combined} and, correspondingly, \eqref{hybrid_combined}, because the ensemble mean has been removed from each of the $N + 1$ ensembles of anomalies in \eqref{hybrid_combined}.
Therefore, the combined ensemble has $N + 1$ degrees of freedom removed instead of the usual one (two for a single static ensemble).
Formally, this choice is immaterial to \eqref{hybrid_covariance}; however, it must match the factor used in standardisations \eqref{s}, \eqref{S}, where it is applied: in EnKF-C $\sqrt{m - 1}$ in \eqref{s} and \eqref{S} is replaced by $\sqrt{m - N - 1}$.
In the EnKF case ($N = 0$) all the above expressions reduce to the usual ones.

The EnOI corresponds to the case $m\subtxt{dyn} = 1$: the single dynamic member is the background $\mb x^f$, there are no dynamic anomalies ($k\subtxt{dyn} = 0$), and $m - N - 1 = \sum_{s=1}^N (m_s - 1)$.
For a single static ensemble with $\gamma_1 = 1$ this yields the ``plain'' EnOI with $\mb P^f = \mb P\suptxt{stat}_1$.

The combined forecast ensemble anomalies \eqref{hybrid_combined} have a larger ensemble size than that of the dynamic ensemble; hence there arises a problem of ensemble reduction to obtain the analysed dynamic ensemble of size $m\subtxt{dyn}$.
EnKF-C takes the approach described in Section~\ref{sec:reduction}, leading to \eqref{reduction_right}:
\begin{align}
  \label{hybrid_update}
  (\mb E\suptxt{dyn})^a = (\mb x^f + \mb A^f \mb w) \mb 1\T + \left(\frac{m\subtxt{dyn} - 1}{m - N - 1}\right)^{1/2} \mb A^f \mb T(:, 1 : m\subtxt{dyn}),
\end{align}
where $\mb x^f$ is the ensemble mean of the dynamic forecast ensemble, $\mb w$ is a vector of ensemble mean update coefficients, and $\mb T$ is the right-multiplied ensemble transform matrix; both $\mb w$ and $\mb T$ are computed using the expanded ensemble anomalies \eqref{hybrid_combined}.
In the EnOI case only the first term of \eqref{hybrid_update} is used.

\chapter{EnKF-C}
\label{ch:enkf-c}

\section{Design considerations}

EnKF-C is designed to use horizontal localisation only.
While some may argue that using vertical localisation might help to decrease the ensemble size, we believe that, for example, in the ocean the vertical structure is too complicated and non-uniform to allow simple and robust solutions in this regard.
Generally, dynamical processes include a variety of barotropic and baroclinic components, and introducing vertical localisation in one form or another can be detrimental for the model's balances.
On the other hand, with an ensemble size of about 100, normally one can ignore the problem of spurious vertical correlations and leave the system to deal with the vertical covariances on its own.

With the system using horizontal localisation only, the model state effectively becomes a collection of independent horizontal fields updated based on their correlations with local ensemble observations.
The assimilation is conducted by calculating a common horizontal array of local ensemble transforms and applying them to each horizontal field of the model.
The local transforms are independent of each other and can be calculated in parallel, as well as the updates of the ensembles of horizontal model fields.

\section{The workflow}

EnKF-C conducts data assimilation in three stages: PREP, CALC, and UPDATE, by running executables \verb|enkf_prep|, \verb|enkf_calc|, and \verb|enkf_update|, correspondingly.
EnKF-C also provides \verb|ens_diag| for calculating a number of ensemble diagnostic variables.

PREP preprocesses observations so that they are ready for DA.
It has the following steps:
\begin{itemize}
\item read original observations and convert them into a vector of structure \verb|observation|;
\item collate them into superobservations;
\item write superobservations to \verb|observations.nc|.
\end{itemize}

PREP does not need to access the model state; it only accesses the model grid.
(Note that for some types of vertical coordinates the vertical model grid may depend on the state.)

CALC calculates ensemble transforms for updating the forecast ensemble of model states (EnKF, Hybrid) or the background model state (EnOI) in the following steps:
\begin{itemize}
\item read superobservations from \verb|observations.nc|;
\item calculate ensemble of forecast observations $\mb H \mb E^f$ (EnKF) or ensemble of background observation anomalies $\mb H \mb A^f$ and background observations $\mb H \mb x^f$ (EnOI);
\item for nodes with specified stride on each horizontal grid get local observations and calculate local ensemble mean update coefficients $\mb w$ and (if the dynamic ensemble size $m\subtxt{dyn} > 1$) local ensemble anomalies transforms $\mb T$ (Sec.~\ref{sec:numerical});
\item save these transforms to \verb|transforms.nc| (or \verb|transforms.nc-0|, \verb|transforms.nc-1|, ... in multi-grid case);
\item calculate and report forecast and analysis innovation statistics;
\item calculate observation impact metrics DFS and SRF (Sec.~\ref{sec:impact}) and save them to \verb|enkf_diag.nc| (or \verb|enkf_diag-<grid ID>.nc| in multi-grid case);
\item at specified horizontal locations save the model state ensemble, observations, transforms/weights, and DA settings to pointlog files (Sec.~\ref{sec:pointlogs}).
\end{itemize}
Apart from this main mode of operation, CALC can also be used for calculating forecast innovations, or operate in the single observation experiment mode.

UPDATE updates the ensemble (EnKF or Hybrid) or the background (EnOI) using the transforms calculated by CALC, along with a number of specified diagnostics, such as the ensemble spread and inflation.

\begin{figure}
  \centering
  \includegraphics[width = 0.99\textwidth]{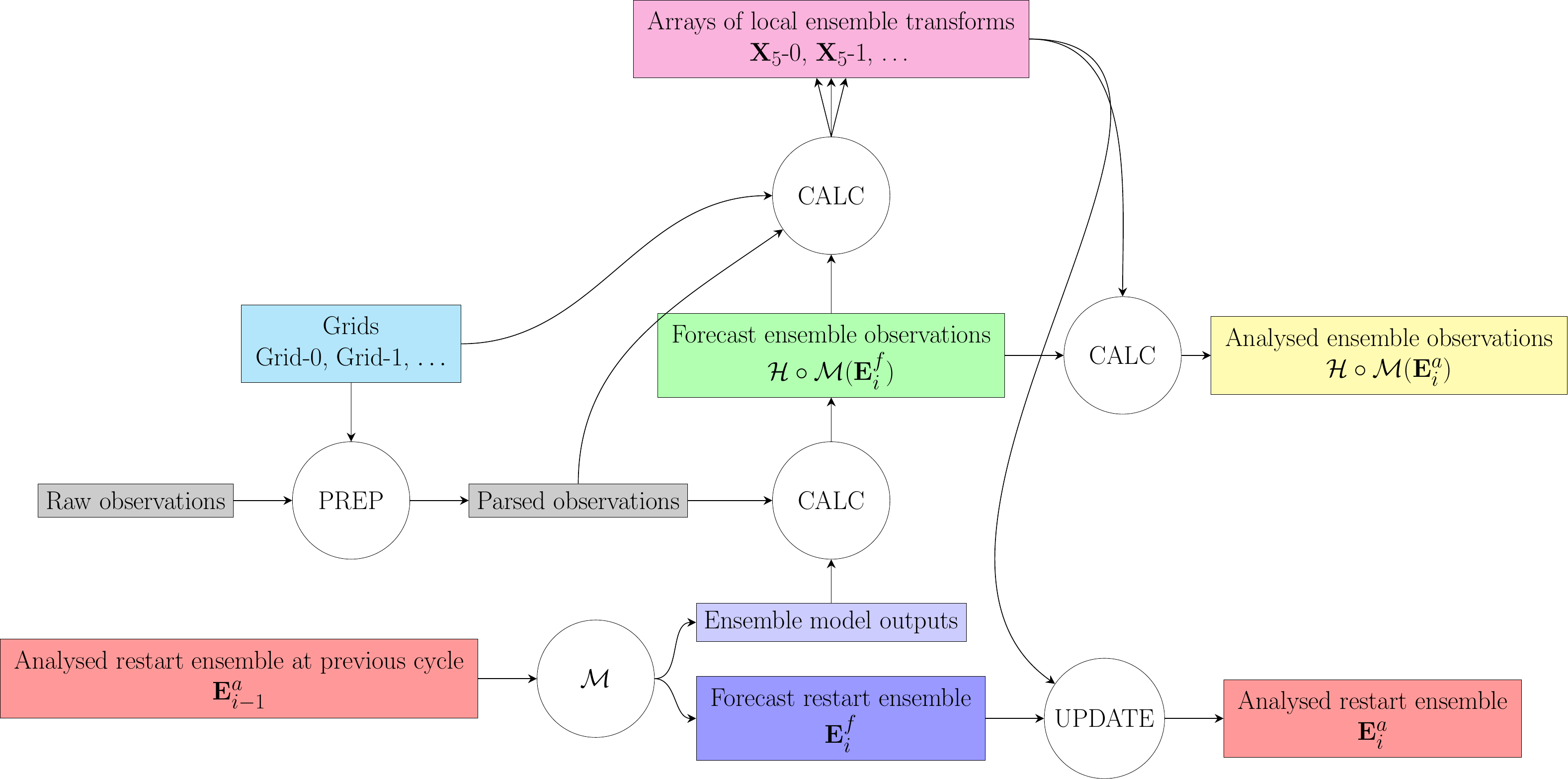}
  \caption{The principle diagram of EnKF-C workflow in EnKF mode.}
  \label{fig:calcupdate}
\end{figure}

The principle diagram of EnKF-C workflow in the EnKF mode is shown in Fig.~\ref{fig:calcupdate}.

\section{Starting up: example 1}
\label{example1}

You may find it helpful to start getting familiar with the system by running the example in \verb|examples/1|.
The example has been put up based on runs of the regional EnKF and EnOI reanalysis systems for Tasman Sea developed by Bureau of Meteorology. 
It allows one to conduct a single assimilation for 23 December 2007 (day 6565 since 1 January 1990) in either EnKF, EnOI, or Hybrid mode.
To reduce the size of the system, the model state has been stripped down to two vertical levels and $100 \times 100$ horizontal grid.
Due to its size (almost 80\,MB) the data for this example is available for download separately from the EnKF-C code -- see \verb|examples/1/README| for details.

\section{Modes of operation}

EnKF-C features three main operation modes: EnKF, EnOI, and Hybrid.
Originally, the system only included the EnKF and EnOI modes, with the EnOI mode handled by a separate codebase.
Later the EnKF mode was extended into the Hybrid mode.
Starting from version 2.48.0, the separate codebase for EnOI was eliminated; both EnKF and EnOI modes are now just particular cases of the Hybrid mode.

The EnKF mode assumes that the size of the static ensemble is zero, and the EnOI mode assumes that the size of the dynamic ensemble is one.
Note that while numerically a Hybrid system with a single dynamic member is equivalent to an EnOI system with the same static ensemble and background, their naming conventions differ:
\begin{itemize}
\item{EnOI System: The background directory is specified by \verb|BGDIR| entry in the main parameter file, and the model files must use the \verb|bg_| prefix.}
\item{Hybrid System: The dynamic member directory is specified by the \verb|ENSDIR_DYNAMIC| entry, and the model files must use the \verb|mem001_| prefix.}
\end{itemize}

\section{Parameter files}

\begin{figure}[h]
  \centering
  \includegraphics[width=0.8\textwidth]{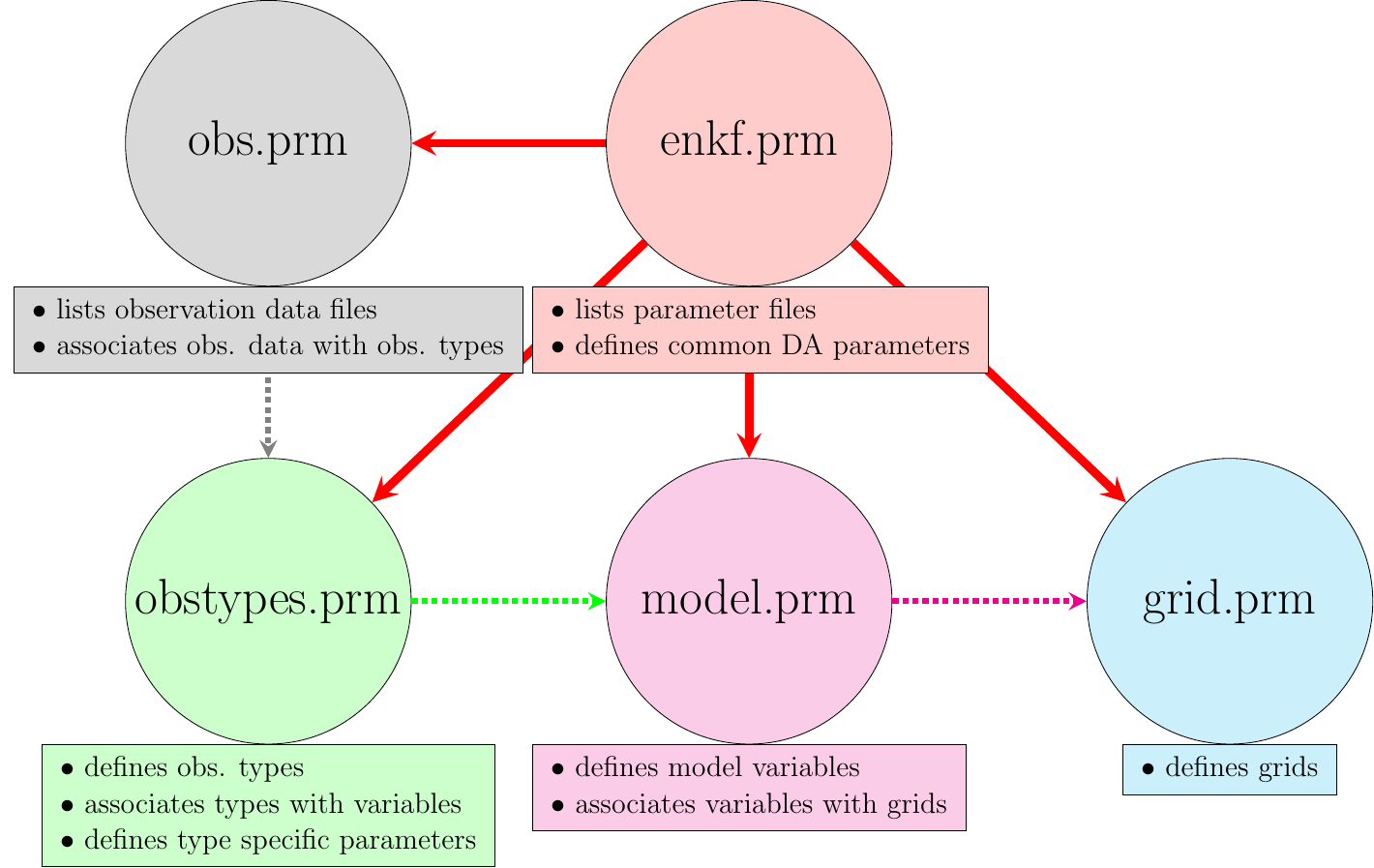}
  \caption{Parameter files in EnKF-C.}
  \label{fig:prm}
\end{figure}

EnKF-C requires 5 parameter files to run (fig.~\ref{fig:prm}):
\begin{itemize} 
\item main parameter file;
\item model parameter file;
\item grid parameter file;
\item observation types parameter file;
\item and observation data parameter file.
\end{itemize}
Examples of these parameter files can be found in \verb|examples/1|.
Running EnKF-C binaries with \verb|--describe-prm-format| in the command line provides information on the parameter file formats.

\subsection{Main parameter file}
\label{sec:mainprm}

The main parameter file specifies the main parameters of DA and four other parameter files.
Its format is described by running \verb|enkf_prep|, \verb|enkf_calc| or \verb|enkf_update| with option \verb|--describe-prm-format|:
\begin{Verbatim}[frame=single,fontsize=\footnotesize]
>./bin/enkf_prep --describe-prm-format

  Main parameter file format:

    TIME            = {<#> days since <YYYY-MM-DD> | <value>}
  [ WINDOWMIN       = <start of obs window in days from analysis> ] (-inf*)
  [ WINDOWMAX       = <end of obs window in days from analysis> ]   (+inf*)
    MODE            = { ENKF | ENOI | HYBRID }
    MODEL           = <model prm file>
    GRID            = <grid prm file>
    OBSTYPES        = <obs. types prm file>
    OBS             = <obs. data prm file>
  (for MODE = ENKF:)
    [ SCHEME                        = { DENKF* | ETKF } ]
    [ ALPHA                         = <alpha> ]                 (1*)
      { ENSDIR | ENSDIR_DYNAMIC }   = <ensemble directory>
    [ { ENSSIZE | ENSSIZE_DYNAMIC } = <ensemble size> ]         (<full>*)
  (for MODE = ENOI:)
      { BGDIR | ENSDIR_DYNAMIC }    = <background directory>

      { ENSDIR | ENSDIR_STATIC }    = <ensemble directory>
    [ { ENSSIZE | ENSSIZE_STATIC }  = <ensemble size> ]         (<full>*)
    [ GAMMA                         = <gamma> ]                 (<1*>)

    [ < more of static ensemble blocks > ]

  (for MODE = HYBRID:)
    [ SCHEME                        = { DENKF* | ETKF } ]
    [ ALPHA                         = <alpha> ]                 (1*)
      ENSDIR_DYNAMIC                = <dynamic ensemble directory>
    [ ENSSIZE_DYNAMIC               = <dynamic ensemble size> ] (<full>*)

      ENSDIR_STATIC                 = <static ensemble directory>
    [ ENSSIZE_STATIC                = <static ensemble size> ]  (<full>*)
      GAMMA                         = <gamma>

    [ < more of static ensemble blocks > ]

  [ KFACTOR         = <kfactor> ]                               (NaN*)
  [ RFACTOR         = <rfactor> ]                               (1*)
    LOCRAD          = <loc. radius in km> ...
    LOCWEIGHT       = <loc. weight> ...                         (# LOCRAD > 1)
  [ NLOBSMAX        = <max. number of local obs. of each type> ]
  [ STRIDE          = <stride for ensemble transforms> ]        (1*)
  [ SOBSTRIDE       = <stride for superobing> ]                 (1*)
  [ FIELDBUFFERSIZE = <field buffer size> ]                     (1*)
  [ FIELDSPLIT      = <field split factor> ]                    (1*)
  [ INFLATION       = <inflation> [ <VALUE>* | PLAIN ]          (1*)
  [ REGION          = <name> <lon1> <lon2> <lat1> <lat2>
    ...
  [ POINTLOG        <lon> <lat> [grid name]]
    ...
  [ EXITACTION      = { BACKTRACE* | SEGFAULT } ]
  [ BADBATCHES      = <obstype> <max. bias> <max. mad> <min # obs.> ]
    ...
  [ ZSTATINTS       = [<z1> <z2>] ... ]
  [ NCFORMAT        = { CLASSIC | 64BIT | NETCDF4 } ]           (NETCDF4*)
  [ NCCOMPRESSION   = <compression level> ]                     (0*)

  Notes:
    1. { ... | ... | ... } denotes the list of possible choices
    2. [ ... ] denotes an optional input
    3. ( ... ) is a note
    4. * denotes the default value
    5. < ... > denotes a description of an entry
    6. ... denotes repeating the previous item an arbitrary number of times
    7. TIME entry is also a flag for geophysical/non-geophysical system,
       depending on the presence of "days since"
\end{Verbatim}
The \verb|TIME| entry also serves as a switch between geophysical and non-geophysical systems.
For geophysical systems the coordinates are assumed to be geographic longitude and latitude (unless \verb|GEOGRAPHIC = 0| is specified for a grid); the time is assumed to be dimensional, in some physical units; and the localisation radii is assumed to be in kilometres.
For non-geophysical systems all the above are assumed to be non-dimensional, and the grid is assumed to be on a plane.

Some further details on the entries of the main parameter file:
\begin{itemize}
\item \verb|DATE| is accepted as a synonym of \verb|TIME| (and is used in the examples).
\item \verb|MODE| must be specified before \verb|ENSDIR|, \verb|ENSSIZE| and \verb|BGDIR|, because the meaning of these entries depends on the mode.
\item Each static ensemble is described by a block starting with \verb|ENSDIR_STATIC| (or \verb|ENSDIR| in the EnOI mode); the entries \verb|ENSSIZE_STATIC| (\verb|ENSSIZE| in the EnOI mode) and \verb|GAMMA| apply to the most recently specified static ensemble directory.
  \verb|GAMMA| defaults to 1 in both EnOI and Hybrid modes (although the describe output above lists it as mandatory for the Hybrid mode) and must be positive.
\item \verb|ENSSIZE_DYNAMIC| may only be set to 1 in the EnOI mode; \verb|SCHEME| and \verb|ALPHA| have no effect when the dynamic ensemble size is 1.
\item \verb|ALPHA| must be in the interval $[0, 1]$.
\item \verb|STRIDE|, \verb|SOBSTRIDE|, \verb|FIELDBUFFERSIZE| and \verb|FIELDSPLIT| must be $\ge 1$; \verb|FIELDBUFFERSIZE| $> 1$ and \verb|FIELDSPLIT| $> 1$ are mutually exclusive.
\item \verb|INFLATION = <value>| is equivalent to \verb|INFLATION = <value> 1|, i.e. the default capping ratio is 1 (Sec.~\ref{sec:capping}).
\item In \verb|REGION| names underscores are replaced by spaces. If no region is specified, a single region ``Global'' is defined.
\item \verb|BADBATCHES| takes exactly four tokens (Sec.~\ref{sec:batches}).
\item \verb|ZSTATINTS| defaults to the intervals $[0, 50]$, $[50, 500]$ and $[500, 9999]$; an empty entry (\verb|ZSTATINTS =|) means no depth intervals. Entry \verb|ZSTATINTS| in the grid parameter file overrides that in the main parameter file.
\item \verb|NCCOMPRESSION| must be between 0 and 9 and requires \verb|NCFORMAT = NETCDF4|.
\end{itemize}

\subsubsection{Global analysis}

It is possible to conduct global analysis by setting \verb|LOCRAD| and \verb|STRIDE| to large numbers.
This is demonstrated by target ``global'' in example 1.

\subsection{Model parameter file}
\label{sec:modelprm}

The model parameter file describes the composition of the state vector by listing the model variables and specifying the associated grids.

\begin{Verbatim}[frame=single,fontsize=\footnotesize]
>./bin/enkf_prep --describe-prm-format model

  Model parameter file format:

    NAME      = <name>
    VAR       = <name>
  [ GRID      = <name> ]                    (# grids > 1)
  [ INFLATION = <value> [<value> | PLAIN] ]
  [ APPLYLOG  = <YES | NO*> ]
  [ RANDOMISE <deflation> <sigma> ]

  [ <more of the above blocks> ]
\end{Verbatim}

Each model variable is described in a block started by the entry for the variable name.
The inflation parameters for a variable, if specified, override the common values set in the main parameter file (Sec.~\ref{sec:capping}).
The convention is the same as in the main parameter file: \verb|INFLATION = <value>| means capped inflation with capping ratio 1, \verb|INFLATION = <value> <ratio>| -- capped inflation with the specified ratio, and \verb|INFLATION = <value> PLAIN| -- plain (uncapped) inflation.
The option APPLYLOG (values \verb|1|, \verb|y|, \verb|yes|, \verb|t| or \verb|true| switch it on) makes it possible to conduct assimilation in log space (Sec.~\ref{sec:logspace}), and RANDOMISE (or RANDOMIZE) -- to specify parameters of ``forgetting'' model for the variable (Sec.~\ref{sec:bias}); the deflation must be between 0 and 1.
Note that any entry starting with ``VAR'' is interpreted as \verb|VAR|, and that grid names are case-sensitive.

EnKF-C permits using multiple model grids.
Each model variable must be associated with one of the grids defined in the grid parameter file.
See \verb|examples/4| for an example.

\subsection{Grid parameter file}
\label{sec:gridprm}

Grid parameter file describes grids used for model variables.
Each grid is described in a section started by the grid name entry and contains the grid name, grid data file, and names of the dimensions and coordinates in the grid data file.
It also contains variable names for the depth and for number of layers in a vertical column (z grids) or land mask (sigma grids):
\begin{Verbatim}[frame=single,fontsize=\footnotesize]
>./bin/enkf_prep --describe-prm-format grid

  Grid parameter file format:

    NAME             = <name> [ PREP | CALC ]
  [ DOMAIN           = <domain name> ]
    DATA             = <data file name>
    (either)
    [ HTYPE          = { rect | curv | unstr | none } ]
      XVARNAME       = <X variable name>
      YVARNAME       = <Y variable name>
      (if htype = unstr)
      [ TRIVARNAME    = <triangle vertice IDS variable name> ]
      [ TRINEIVARNAME = <triangle neighbour IDS variable name> ]
      (end if)
    (or)
      HGRIDFROM      = <grid name>
    (end either)
    VTYPE            = { z | zt | sigma | hybrid | numeric | none }
  [ VDIR             = { fromsurf* | tosurf } ]
  [ GEOGRAPHIC       = { 0 | 1* | 2 } ]
    (if vtype = z)
      ZVARNAME       = <Z variable name>
    [ ZCVARNAME      = <ZC variable name> ]
    [ NUMLEVELSVARNAME = <# of levels variable name> ]
    (else if vtype = zt)
      ZVARNAME       = <Z variable name>
    [ NUMLEVELSVARNAME = <# of levels variable name> ]
    (else if vtype = sigma)
      CVARNAME       = <Cs_rho variable name>
    [ CCVARNAME      = <Cs_w variable name> ]
    [ SVARNAME       = <s_rho variable name> ]            (uniform*)
    [ SCVARNAME      = <s_w variable name> ]              (uniform*)
    [ HCVARNAME      = <hc variable name> ]               (0.0*)
    [ MASKVARNAME    = <land mask variable name> ]
    (else if vtype = hybrid)
      AVARNAME       = <A variable name>
      BVARNAME       = <B variable name>
    [ ACVARNAME      = <AC variable name> ]
    [ BCVARNAME      = <BC variable name> ]
      P1VARNAME      = <P1 variable name>
      P2VARNAME      = <P2 variable name>
    [ MASKVARNAME    = <land mask variable name> ]
    (else if vtype = numeric)
      ZVARNAME       = <Z variable name>
    [ ZCVARNAME      = <ZC variable name> ]
    [ NUMLEVELSVARNAME = <# of levels variable name> ]
    (end if)
  [ DEPTHVARNAME   = <depth variable name> ]
  [ STRIDE           = <stride for ensemble transforms> ] (1*)
  [ SOBSTRIDE        = <stride for superobing> ]          (1*)
  [ ZSTATINTS        = [<z1> <z2>] ... ]

  [ <more of the above blocks> ]
\end{Verbatim}

While the code is supposed to automatically identify the types of horizontal grids used, they can also be specified explicitly.
The following (case-insensitive) values of \verb|HTYPE| are accepted: \verb|rect| or \verb|rectangular|; \verb|curv| or \verb|curvilinear|; \verb|curv2| or \verb|curvilinear2| (curvilinear grid rendered using stereographic projections, chosen automatically for curvilinear grids with \verb|GEOGRAPHIC = 2|); \verb|unstr|, \verb|unstruct| or \verb|unstructured|; and \verb|none|.
The type of the vertical grid has to be specified, and the entry \verb|VTYPE| must precede the entries describing the vertical coordinate.
If some grid (say, grid A) has the same horizontal grid as another grid (grid B), the code can be notified of this by entering name of grid A in the field \verb|HGRIDFROM| for grid B (or vice versa).
This saves memory, grid initialisation time, and uses transforms calculated for grid A for all variables defined on grid B.

Further details:
\begin{itemize}
\item A qualifier \verb|PREP| or \verb|CALC| after the grid name (\verb|NAME = <name> PREP|) makes the grid visible only to PREP or only to CALC and UPDATE, respectively. This makes it possible, for example, to use simplified grid specifications in PREP.
\item The entry for the number of levels / land mask depends on the vertical grid type: it is \verb|NUMLEVELSVARNAME| for ``z'', ``zt'' and ``numeric'' grids and \verb|MASKVARNAME| for ``sigma'', ``hybrid'' and ``none'' grids. This entry can be followed by qualifiers \verb|PREP|, \verb|CALC| and/or \verb|UPDATE| to restrict it to particular programs.
\item \verb|DEPTHVARNAME| can be specified for grids of any vertical type.
\item \verb|STRIDE| for a grid defaults to the \verb|STRIDE| in the main parameter file; on unstructured grids it is always 1.
\item \verb|SOBSTRIDE| in the grid parameter file is read but currently ignored: the superobing stride is defined by \verb|SOBSTRIDE| in the observation types parameter file or, by default, in the main parameter file.
\item \verb|ZSTATINTS| for a grid overrides that in the main parameter file (Sec.~\ref{sec:mainprm}).
\end{itemize}

\subsubsection{Horizontal grids}
\label{sec:hgrid}

At the moment, EnKF-C supports 3 main types of horizontal grids:
\begin{itemize}
\item rectangular quadrilateral grids aligned with geographic or Cartesian coordinates;
\item curvilinear quadrilateral grids;
\item unstructured grids.
\end{itemize}
A grid is assumed to be rectangular if grid node coordinates depend on one dimension, and the coordinate dimensions differ for X and Y coordinates.
A grid is assumed to be curvilinear if the grid node coordinates depend on two dimensions.
A grid is assumed to be unstructured if grid node coordinates depend on the same dimension for X and Y coordinates.
For rectangular grids the code tries to determine and handle periodicity in X direction.

Note that the code does not detect and therefore cannot take advantage of periodic curvilinear grids; because of that, it does not map (skips) observations in cells connecting the grid edges.

The flag \verb|GEOGRAPHIC| was introduced in v2.8.0.
When set to ``1'' (``yes'', ``true''), it communicates to the code that the grid nodes are defined in geographic coordinates.
In that case the grid is rendered using 3D Cartesian coordinates; otherwise it is rendered in 2D geographic or Cartesian coordinates, which is somewhat less expensive.
The more strict 3D rendering can be essential in polar regions because of the skewness of grid cells in geographic coordinates there, which can cause failures of the mapping algorithm.
One can investigate the necessity of setting this flag to ``1'' by comparing the number of accepted/rejected observations with and without it.

If the flag \verb|GEOGRAPHIC| is set to ``0'' then the grid is assumed to be on a plane and grid coordinates are assumed to be non-dimensional; the localisation radii for the corresponding observation types need to be specified accordingly.

The flag \verb|GEOGRAPHIC| set to ``2'' treats grid coordinates similarly to that for ``1'' but switches to more robust rendering procedures (based on stereographic projections) and is advised to be used e.g. for grids with artificial discontinuities.

For unstructured grids associated with observed variables it is necessary to specify triangulation details: vertex IDs (\verb|TRIVARNAME|) and neighbour triangle IDs (\verb|TRINEIVARNAME|).
Both must be specified together, and both variables must be integer arrays of shape $(n_{tri}, 3)$ with 0-based indices (that is, vertex IDs between 0 and $n_{nodes} - 1$, and neighbour IDs between 0 and $n_{tri} - 1$, with a negative value for a missing neighbour); the code checks the type, shape and range of these arrays.
For unstructured grids stride can be equal to 1 only (i.e. ensemble transforms must be calculated in each grid node).

\subsubsection{Vertical grids}

The vertical coordinates are used for mapping the depth/height (pressure) of non-surface observations to fractional layer index.
The observation depth or height are assumed to be positive.

The type of the vertical grid is defined by entry \verb|VTYPE| in the grid parameter file.
EnKF-C supports the following types of vertical grids: ``z'' ($Z$); ``zt''; ``sigma'' ($\sigma$); ``hybrid'' ($\sigma-p$), and ``numeric''.
It is also possible to define a purely horizontal (two-dimensional) grid by defining its vertical type as ``none''.

For $Z$ grids one needs to define vertical coordinates of layer centres (entry \verb|ZVARNAME|) and (optionally) the coordinates of layer boundaries (\verb|ZCVARNAME|).
The layer centres must be strictly monotonic.
If coordinates of layer boundaries are not specified they are built by the code so that the layer centres are in the middle of intervals between boundaries.
The ZC variable can contain either $n_k + 1$ values (all layer boundaries) or $n_k$ values.
In the latter case each value is taken to be the boundary of the corresponding layer on the side \emph{away from the surface} (as in MOM), and the surface boundary is set to $z = 0$.
For backward compatibility, for $Z$ grids with \verb|VDIR = FROMSURF| $n_k$ values are interpreted as surface-side boundaries if the first value is on the surface side of the first layer centre.
\verb|ZCVARNAME| cannot be specified for ``zt'' grids.

For ``zt'' grids the layer boundaries are placed at the midpoints between the adjacent layer centres, assuming that the surface layer starts at $z = 0$ (provided that $z = 0$ is on the surface side of the surface layer centre), and extrapolating the bottom boundary by half of the distance between the two deepest layer centres.
\verb|ZCVARNAME| cannot be specified for ``zt'' grids.

For $\sigma$ grids the code implements the ``new'' vertical coordinate formulation from ROMS as described in \url{https://www.myroms.org/wiki/Vertical_S-coordinate}, Eq.~2 and elaborated by Shchepetkin in \url{https://www.myroms.org/forum/viewtopic.php?f=20&t=2189}.
This formulation reduces to the ``standard'' $\sigma$ grid if the entry \verb|HCVARNAME| is not specified or if the corresponding variable in the grid file is set to zero.
Similarly to $Z$ grid, one needs to define the vertical coordinates of layer centres (entry \verb|CVARNAME|) and (optionally) the coordinates of layer corners (entry \verb|CCVARNAME|).
From version 1.81.0 one can also specify variables for layer coordinates (which is used to be ``plain'' sigma, that is uniform) via entries \verb|SVARNAME| and \verb|SCVARNAME|.

The hybrid $\sigma$-$p$ grids are implemented as described in \url{https://journals.ametsoc.org/doi/pdf/10.1175/2008MWR2537.1}.
One needs to specify the $A$ (\verb|AVARNAME|) and $B$ (\verb|BVARNAME|) arrays for layer centres as well as the top and surface pressure (\verb|P1VARNAME| and \verb|P2VARNAME|).
One can also specify optional $A$ and $B$ arrays for layer corners (\verb|ACVARNAME|, \verb|BCVARNAME|); these must be specified together.
The entry \verb|DEPTHVARNAME| needs only to be specified for variables with non-surface observations.

Vertical grid type ``numeric'' is supposed to handle general situations described by 3D arrays of layer depths.
Grids of this type are defined by specifying variable with cell centres depth coordinates (entry \verb|ZVARNAME|) and (optionally) variable with cell boundaries depth coordinates (entry \verb|ZCVARNAME|).
The ZC variable must have the same horizontal dimensions as the Z variable and either $n_k + 1$ or $n_k$ layers; the latter case is treated in the same way as for $Z$ grids (boundaries on the side of each layer away from the surface).

By default, the code assumes that the surface is at layer 0.
If this is not the case, one needs to describe it explicitly by the entry \verb|VDIR = TOSURF|; otherwise the surface ensemble or background observations will not be calculated correctly.

The vertical interpolation is performed linearly assuming constant gradients within layers.
This means that at the boundary between layers the variables have middle value between those at the adjacent layer centres.
This rule applies to vertical grids of all types but ``zt'', for which vertical interpolation assumes constant gradients between layer centres.

\subsubsection{``Empty'' grids}

``Empty'' grids have been introduced in version 2.28.0 to simplify estimation of global scalar parameters.
In the KF context a model parameter is simply an unobserved model variable.
In EnKF-C model variables are defined in the model parameter file; each model variable must be associated with a grid defined in the grid parameter file.
``Empty'' grids are used to be associated with global model parameters and are defined by specifying their vertical and horizontal types as ``none'' (with no \verb|DATA| entry), e.g.
\begin{Verbatim}[frame=single,fontsize=\footnotesize]
  NAME = param-grid
  VTYPE = none
  HTYPE = none
\end{Verbatim}
Note that like ``normal'' grids, the empty grids can be associated with particular domains (see Sec.~\ref{sec:domains}) to be updated from observations of certain types only.

\subsection{Observation types parameter file}
\label{sec:obstypesprm}

Observation types are the interface that connects model and observations.
They are specified in a separate parameter file.
Each observation type is described in a separate section identified by the entry \verb|NAME|.
Apart from the type name, the section must contain the entries \verb|ISSURFACE| and \verb|VAR| (the associated model variable(s)); the observation operator (\verb|HFUNCTION|) is optional and defaults to ``standard'' (Sec.~\ref{sec:hfunctions}).
The optional parameters include the R-factor and localisation radius for the type (Sec.~\ref{sec:datuning}), the allowed range, and spatial limits for the corresponding observations.
\begin{Verbatim}[frame=single,fontsize=\footnotesize]
>./bin/enkf_prep --describe-prm-format obstypes

  Observation types parameter file format:

    NAME        = <name>
  [ DOMAINS     = <domain name> ... ]
    ISSURFACE   = {yes | no}
  [ STATSONLY   = {yes | no*} ]
    VAR         = <model variable name> ...
  [ ALIAS       = <variable name used in file names> ]   (VAR*)
  [ OFFSET      = <file name> <variable name> ]          (none*)
  [ MLD_VARNAME = <model varname> ]                      (none*)
  [ MLD_THRESH  = <threshold> ]                          (NaN*)
  [ HFUNCTION   = <H function name> ]                    (standard*)
  [ ASYNC       = {<time interval> {centre | endpoint} <time varname>} | {no|false|0} (0*)
  [ LOCRAD      = <loc. radius in km> ... ]
  [ LOCWEIGHT   = <loc. weight> ... ]                    (# LOCRAD > 1)
  [ RFACTOR     = <rfactor> ]                            (1*)
  [ ERROR_DOUBLING_TIME = <time in days> ]               (inf*)
  [ NLOBSMAX    = <max. allowed number of local obs.> ]  (inf*)
  [ ERROR_STD_MIN = <min. allowed superob error> ]       (0*)
  [ SOBSTRIDE   = <stride for superobing> ]              (1*)
  [ PERMIT_LOCATION_BASED_THINNING = {yes* | no} ]
  [ MINVALUE    = <minimal allowed value> ]              (-inf*)
  [ MAXVALUE    = <maximal allowed value> ]              (+inf*)
  [ XMIN        = <minimal allowed X coordinate> ]       (-inf*)
  [ XMAX        = <maximal allowed X coordinate> ]       (+inf*)
  [ YMIN        = <minimal allowed Y coordinate> ]       (-inf*)
  [ YMAX        = <maximal allowed Y coordinate> ]       (+inf*)
  [ ZMIN        = <minimal allowed Z coordinate> ]       (-inf*)
  [ ZMAX        = <maximal allowed Z coordinate> ]       (+inf*)
  [ WINDOWMIN   = <start of obs window in days from analysis> ] (-inf*)
  [ WINDOWMAX   = <end of obs window in days from analysis> ]   (+inf*)
  [ EXCLUDE     = <lon1> <lon2> <lat1> <lat2> | <mask file> <varname>]
    ...
  
  [ <more of the above blocks> ]
  [ EXCLUDE     = <lon1> <lon2> <lat1> <lat2> | <mask file> <varname>]  
\end{Verbatim}

The tags for available observation operators are listed in array \verb|allhentries| in file \verb|calc/allhs.c|.

The entry \verb|ASYNC| (Sec.~\ref{sec:async}) has the form \verb!ASYNC = <dt> {centre | endpoint} <time varname>! (the alignment can be abbreviated to \verb|c| or \verb|e|); there is no default alignment and the time variable name is mandatory.
\verb|ASYNC = no| (or \verb|false|, \verb|0|) sets the type to be synchronous (the default).

The \verb|OFFSET| entry may be used for adding the known model bias to observations, for example, to specify the mean dynamic topography (MDT) when assimilating sea level anomaly (SLA) observations.
The dimension of the offset should match that of the corresponding model variable, except that it is possible to define (1D) layer-wise offsets for 3D model variables.
\verb|OFFSET| cannot be used for observation types associated with log-transformed model variables (Sec.~\ref{sec:logspace}).

The localisation radius (and \verb|NLOBSMAX|, \verb|SOBSTRIDE|) for an observation type, if specified, overrides the common value from the main parameter file.
\verb|SOBSTRIDE| greater than 1 is not allowed for types associated with unstructured grids.
The R-factors for each observation type are obtained by multiplying the common
value by the observation type value.
(More on localisation radius and R-factor in Section~\ref{sec:datuning}.)

The entries \verb|MLD_VARNAME| and \verb|MLD_THRESH| are used to calculate the model mixed layer depth for projecting the surface bias.

\verb|WINDOWMIN| and \verb|WINDOWMAX| define the allowed temporal interval for this observation type relatively to the analysis day and override the corresponding common settings in the main parameter file.

\verb|ERROR_DOUBLING_TIME| $T$ (in days) increases the observation error of each observation of this type according to the time distance from the analysis:
$\sigma_{obs} \leftarrow \sigma_{obs} \, 2^{|t - t_a| / T}$, where $t$ is the observation time and $t_a$ is the analysis time.
It is applied by PREP to the original observations after all \verb|ERROR_STD| entries in the observation data parameter file have been processed.

The allowed observation domain is defined by \verb|XMIN|, \verb|XMAX|, \verb|YMIN|, \verb|YMAX| (observations on these bounds are discarded) and \verb|ZMIN|, \verb|ZMAX| (observations on these bounds are retained).

\verb|DOMAINS| lists the names of grid domains (Sec.~\ref{sec:domains}) the observations of this type are visible from; by default (``ALL'') they are visible from all grids.

It is possible to specify regions with no observations (if, for example, the updated model becomes unstable at some location).
This is done with entries \verb|EXCLUDE|.
The regions are related either to observations of certain type (if defined in
a type section) or to observations of all types (if defined outside any type
section).

The use of \verb|ALIAS| is described in Section~\ref{sec:filenames}.

\subsection{Observation data parameter file}

Observation data parameter file specifies observations to be assimilated.
EnKF-C has a simple policy in this regard: if a data file is listed in the observation data parameter file, then observations from this file are assimilated.
This allows one using custom observation windows for particular observation types, instruments etc., specifying details on the script level during the parameter file generation.

In practice some of observations specified in the observation data parameter file can be outside the observation time window for the cycle.
In this case the exact boundaries of the observation window can be specified by entries \verb|WINDOWMIN| and \verb|WINDOWMAX| in the main parameter file or (for specific observation types) observation types parameter file;
observations with time outside interval \verb|[TIME+WINDOWMIN, TIME+WINDOWMAX)| will be discarded (note that \verb|WINDOWMIN| is normally negative).
The window set for an observation type overrides that in the main parameter file.

The observation parameter file contains an arbitrary number of sections identified by entries \verb|PRODUCT|.
Each section specifies the observation type, input files, reader and, possibly, observation error:
\begin{Verbatim}[frame=single,fontsize=\footnotesize]
./bin/enkf_prep --describe-prm-format obsdata

  Observation data parameter file format:

    PRODUCT   = <product>
    TYPE      = <observation type>
    READER    = <reader>
    FILE      = <data file wildcard> 
    ...
  [ ERROR_STD = { <value> | <data file> <varname> } [ EQ* | PL | MU | MI | MA ] ]
    ...
  [ MANDATORY = {yes | no*} ]
  [ PARAMETER <name> = <value> ]
    ...

  [ <more of the above blocks> ]

  Notes:
    1. { ... | ... | ... } denotes the list of possible choices
    2. [ ... ] denotes an optional input
    3. * denotes the default value
    4. < ... > denotes a description of an entry
    5. ... denotes repeating the previous item an arbitrary number of times
    6. Use "--list-readers" to list available readers
    7. Use "--describe-reader <reader>" to list available parameters
\end{Verbatim}
Observation files can be defined using wildcards ``*'' and ``?''.
A \verb|FILE| entry that does not match any existing file is considered to be a fatal error (since v2.43.11).
If \verb|MANDATORY = yes| is specified, PREP also stops if the section yields no observations.

If \verb|READER| is not specified, the reader ``scattered'' is used.
Reader names are case-sensitive.
Each \verb|PARAMETER| can only be specified once in a section, except for \verb|QCFLAGNAME|, \verb|QCFLAGVALS| and \verb|EXCLUDEINST| (and any parameter for the reader ``profile'').

The line in the above example starting with \verb|ERROR_STD| specifies the observation error.
It can contain either a number or a file name.
In the case of entering the file name there also should be another entry in the same line specifying the name of the variable to be read.
The variable should be a 2D or 3D field on the grid of the observation type, depending on whether the observation type is a surface type or not (entry \verb|ISSURFACE| in the observation types parameter file, Sec.~\ref{sec:obstypesprm}); a 3D field is accepted for a surface type, in which case its surface layer is used.
The field is interpolated to the observation locations.

The line with observation error can also have another token specifying the type of operation to be conducted: \verb|EQUAL| ($\sigma_{tot} \leftarrow \sigma_{now}$, default), \verb|PLUS| ($\sigma_{tot} \leftarrow \sqrt{\sigma_{tot}^2 + \sigma_{now}^2}$), \verb|MULT| ($\sigma_{tot} \leftarrow \sigma_{tot} \sigma_{now}$), \verb|MIN| ($\sigma_{tot} = \max(\sigma_{tot}, \sigma_{now})$), or \verb|MAX| ($\sigma_{tot} = \min(\sigma_{tot}, \sigma_{now})$).
(Only the first two letters of the operation are checked: \verb|EQ|, \verb|PL|, \verb|MU|, \verb|MI|, \verb|MA|.)
There can be several error entries in a section in the observation parameter file; they are applied in the order of their appearance.
Error values must be non-negative.
\verb|ERROR_STD| must be specified explicitly for observation types associated with log-transformed model variables (Sec.~\ref{sec:logspace}).

The observation time only matters if the observation type is specified to be ``asynchronous'' (see Sec. \ref{sec:async}).
In this case the model estimation for the observation is made by using model state at the appropriate time.
Otherwise, observations are assumed to be made at the time of assimilation, regardless of the actual observation time.

Note that there can be multiple blocks with the same product.
This enables custom treatment of some specific data.
For example, the following entries override observation error for Geosat (files with prefix \verb|g1_|) on 23 May 2006:
\begin{Verbatim}[frame=single,fontsize=\footnotesize]
# set observation error for Geosat to 7cm
PRODUCT = RADS
TYPE = SLA
READER = scattered
PARAMETER VARNAME = sla
PARAMETER ZVALUE = NaN
PARAMETER MINDEPTH = 100
FILE = /short/p93/pxs599/obs/RADS/2006/g?_20060523.nc
ERROR_STD = 0.07

# use default errors for other altimeters
PRODUCT = RADS
TYPE = SLA
READER = scattered
PARAMETER VARNAME = sla
PARAMETER ZVALUE = NaN
PARAMETER MINDEPTH = 100
file = /short/p93/pxs599/obs/RADS/2006/[!g]?_20060523.nc
\end{Verbatim}

\section{File name conventions}
\label{sec:filenames}

EnKF-C assumes that the ensemble and background file names have some predefined formats.
The file name \verb|fname| for ensemble member \verb|memberid| containing model variable \verb|varname| is assumed to be \spverb|sprintf(fname, "mem
This file typically represents the model restart (EnKF, dynamic members) or ensemble anomaly (static members).

For the EnOI the background (forecast restart) file for variable \verb|varname| is assumed to be \spverb|sprintf("bg_

The analysis restart file names are created by concatenating the above forecast restart names and a suffix, either``.analysis'' or ``.increment'', depending on the output of UPDATE.

For asynchronous DA the model output file names for the time slot \verb|t| are assumed to be \\\spverb|sprintf("mem
It is also possible to have model outputs concatenated into a single multi-record file (or several such files), in which case their names are assumed to match either \spverb|sprintf("mem
In this case EnKF-C will identify the record to be used by its time.
Asynchronous model outputs are looked up in the dynamic ensemble (background) directory only; static members are always read synchronously from \spverb|ENSDIR_STATIC/mem
(See Sec.~\ref{sec:async}.)

There are possible situations when the surface field and 3D field of the same variable have different asynchronous settings.
For example, the sea surface temperature (SST) may have asynchronous time intervals of 0.25 days, while for the subsurface temperature these may be set to 1 day.
In such cases there is a clash between the corresponding asynchronous file names.
To resolve it, one (or both) fields should use an alias instead of the model variable name in its file name specified by the entry ALIAS of the corresponding observation type.

\section{PREP}

PREP is the first stage of data assimilation in EnKF-C.
It preprocesses observations by bringing them to a common form and merging close observations into so called superobservations.

By design, PREP is supposed to be light-weight, so that it does not read either the ensemble or background, and the only model information it needs is the model grid. 
(Note that this may require some additional processing at later stages for models with dynamic grid, such as HYCOM.)

The name of the binary (executable) for PREP is \verb|enkf_prep|.
It has the following usage and options:
\begin{Verbatim}[frame=single,fontsize=\footnotesize]
>./bin/enkf_prep
  Usage: enkf_prep <prm file> [<options>]
  Options:
  --consider-subgrid-variability
      increase error of superobservations according to subgrid variability
  --describe-prm-format [main|model|grid|obstypes|obsdata]
      describe format of a parameter file and exit
  --describe-reader <reader>
      decribe reader and exit
  --describe-superob <sob #>
      print composition of this superobservation and exit
  --list-readers
      list available readers and exit
  --no-superobing
  --no-superobing-across-batches
  --no-thinning
  --superob-across-instruments
  --write-orig-obs
      write original obs within model domain to observations-orig.nc
  --write-all-orig-obs
      write all original obs to observations-orig.nc
  --version
      print version and exit
\end{Verbatim}

\verb|enkf_prep| writes the preprocessed observations to file \verb|observations.nc|.
When run with command line argument \verb|--write-orig-obs|, it also writes the original (not superobed) observations to \verb|observations-orig.nc|.
By default, the original observations only involve observations within the corresponding model grids, but can include all observations by the command line argument \verb|--write-all-orig-obs|.
In \verb|observations-orig.nc| the auxiliary variable \verb|aux| contains the ID of the superobservation in \verb|observations.nc| each original observation has been collated into ($-1$ for observations that did not make it into a superobservation).

\subsection{Observation types, products, instruments, batches, readers}

\subsubsection{Types}
\label{sec:types}

Each observation has a number of attributes defined by the fields of the structure \verb|observation|.
One of them is observation type, which characterises the observation in a general way and relates it to the model state.
For example, typical oceanographic observations may have tags SLA (for sea level anomalies), SST (sea surface temperature), TEM (subsurface temperature) and SAL (subsurface salinity).
Different types can be related to the same model variable, as do SST and TEM in the above example.
Observation types are described in the corresponding parameter file (Sec.~\ref{sec:obstypesprm}).

\subsubsection{Products}

An observation is also characterised by ``product''.
It can be a tag for the data set, e.g.:
\begin{Verbatim}[frame=single,fontsize=\footnotesize]
PRODUCT = RADS
TYPE = SLA
READER = scattered
PARAMETER VARNAME = sla
PARAMETER BATCHNAME = pass
PARAMETER ZVALUE = NaN
PARAMETER MINDEPTH = 100
FILE = obs/RADS/2007/??_200712{19,20,21,22,23}.nc

PRODUCT = ESACCI
TYPE = SST
READER = scattered
PARAMETER VARNAME = sst
PARAMETER ZVALUE = 0
PARAMETER VARSHIFT = -273.15
FILE = obs/ESACCI/2007/200712{19,20,21,22,23}-*.nc
\end{Verbatim}

\subsubsection{Instruments}

The observational data in a product can be collected by a number of instruments.
The corresponding field in the \verb|measurement| structure is supposed to be filled by the observation reader.

\subsubsection{Batches}
\label{sec:batches}

An observation can be attributed to one of the groups called ``batches'', such as altimeter passes, Argo profiles etc., to enable detection and discarding of bad batches.
A unique batch of observations is defined by (1) observation type; (2) observation data file; and (3) ``batch'' identifier set by the observation reader.
With generic readers the batch identifier can be set to a variable in the data file via parameter \verb|BATCHNAME|; the custom readers set it internally.

A batch of observations is considered bad if either its mean innovation or mean absolute innovation exceed specified thresholds.
Specifications for bad batches can be set in the parameter file as follows:
\begin{Verbatim}[frame=single,fontsize=\footnotesize]
BADBATCHES = SLA 0.06 0.10 500
BADBATCHES = TEM 4 5 0
BADBATCHES = SST 0.3 0.5 10000
BADBATCHES = SAL 1.5 2 0
\end{Verbatim}
The above entry means that any batch of observations of type SLA (typically, an orbit) containing at least 500 observations and having either mean innovation of at least 0.06 (meter) in magnitude or mean absolute innovation of at least 0.10 is considered to be bad.
Similarly, a TEM batch (typically, a profile) is considered bad if the mean innovation exceeds 4 (degrees) or the mean absolute innovation exceeds 5 (degrees).
The parameter file can have an arbitrary number of such entries.

The bad batches are identified and removed from further processing by CALC.
From version 2.15.0 this no longer requires two passes of PREP and CALC.
Information about all observation batches is written by \verb|enkf_calc| to the file \verb|obsbatches.txt|, and about the detected bad batches -- to the file \verb|badobsbatches.txt|.
Note that for observations from bad batches the value of variable \verb|status| in \verb|observations.nc| is set by CALC to \verb|STATUS_BADBATCH| (currently 5).

\subsubsection{Readers}

The function of data readers is to read observations in specified files and parse them sequentially into \verb|struct observation| defined in \verb|common/observations.h|.
Following is the list of available readers:
\begin{Verbatim}
>bin/enkf_prep --list-readers
  generic readers:
    scattered
    gridded_xy
    gridded_xyz
    gridded_xyh
    profile
  custom readers:
    navo
    windsat
    mmt
    amsr2
    amsre
    cmems
    en4
\end{Verbatim}

Users are encouraged to use generic rather than custom readers.
When this is not possible, one may either modify (or put a request to modify) a generic reader, or develop a custom reader.
Note that reader names are case-sensitive, and that each custom reader can only be used with its own product (\verb|NAVO|, \verb|WINDSAT|, \verb|MMT|, \verb|AMSR2|, \verb|AMSRE|, \verb|CMEMS| or \verb|EN4|, matched case-insensitively), while generic readers can be used with any product.
The reader ``profile'' can also be referred to as ``z''.

Each reader can be specified in the observation data parameter file with an arbitrary number of parameters.
For example, the following section sets the default minimal depth for SLA observations to 150\,m:
\begin{Verbatim}
(...)
PRODUCT = RADS
TYPE = SLA
READER = scattered
PARAMETER VARNAME = sla
PARAMETER ZVALUE = NaN
PARAMETER MINDEPTH = 150
(...)
\end{Verbatim}

Observation data parameters can be either generic (common for all readers), or custom (specific to specific readers or groups of readers).
The generic parameters include:
\begin{description}
\item{\verb|MINDEPTH|} -- minimal allowed \emph{model} depth at the observation location (requires \verb|DEPTHVARNAME| for the grid);
\item{\verb|MAXDEPTH|} -- maximal allowed model depth (requires \verb|DEPTHVARNAME| for the grid);
\item{\verb|FOOTPRINT|} -- the radius in km of the horizontal footprint;
\item{\verb|VARSCALE|} -- data scale;
\item{\verb|VARSHIFT|} -- data offset;
\item{\verb|STRIDE|} (or \verb|THIN|) -- stride interval: only every \verb|STRIDE|-th observation read in this section is retained;
\item{\verb|YMIN|} -- minimal allowed latitude (exclusive);
\item{\verb|YMAX|} -- maximal allowed latitude (exclusive);
\item{\verb|ZMIN|} -- minimal allowed \emph{observation} depth (inclusive);
\item{\verb|ZMAX|} -- maximal allowed observation depth (inclusive);
\item{\verb|LOCATION_BASED_THINNING_TYPE|} -- type of location based thinning: \verb|XYZ| (default), \verb|XY|, \verb|CELL|, or \verb|0| (\verb|no|, \verb|false|) for no thinning (Sec.~\ref{sec:superobing}).
\end{description}
Specified parameters apply to observations defined in the corresponding section of observation data parameter file.
These parameters can be used with any reader, including custom readers.

The custom parameters can be learned by running \verb|enkf_prep --describe-reader <reader name>|.
Following is the description of the reader \verb|scattered|:
\begin{Verbatim}
>./bin/enkf_prep --describe-reader scattered

  Generic reader "scattered" reads 2D or 3D scattered point data.

  There are a number of parameters that must (marked below with "+") or may ("-")
  be specified in the corresponding section of the observation data parameter file.
  For non-specified parameters a list of default variable names will be checked and
  the first matching variable will be used as the paramater value.

  Each parameter needs to be entered as follows:
    PARAMETER <name> = <value> ...

  Parameters common to generic readers:
    - VARNAME (+)
    - TIMENAME (+) ("t" | "[tT]ime" | "TIME")
    - or TIMENAMES (+) (when time = base_time + offset)
    - LONNAME (+) ("lon" | "[lL]ongitude" | "LONGITUDE")
    - LATNAME (+) ("lat" | "[lL]atitude" | "LATITUDE")
    - ZNAME (+) ("z" | "[dD]epth" | "DEPTH") | ZVALUE (+)
        "ZNAME" is needed for 3D data, "ZVALUE" for 2D data (can be NaN)
    - STDNAME (-) ("std")
        dispersion of the collated data
    - ESTDNAME (-) ("error_std")
        error STD; if absent then needs to be specified in the corresponding
        section of the observation data parameter file
    - BATCHNAME (-) ("batch")
        name of the variable used for batch ID (e.g. "pass" for SLA)
    - INSTRUMENT (-)
        instrument string that will be used for calculating instrument stats
        (overrides the global attribute "instrument" in the data file)
    - INSTATTNAME (-)
        name of the global attribute for the instrument string
    - INSTPREFIX (-)
        prefix that will be added in front of the instrument string
    - QCFLAGNAME (-)
        name of the QC flag variable, possible values 0 <= qcflag <= 31
    - QCFLAGVALS (-)
        the list of allowed values of QC flag variable
        Note: it is possible to have multiple entries of QCFLAGNAME and
        QCFLAGVALS combination, e.g.:
          PARAMETER QCFLAGNAME = TEMP_quality_control
          PARAMETER QCFLAGVALS = 1
          PARAMETER QCFLAGNAME = DEPTH_quality_control
          PARAMETER QCFLAGVALS = 1
          PARAMETER QCFLAGNAME = LONGITUDE_quality_control
          PARAMETER QCFLAGVALS = 1,8
          PARAMETER QCFLAGNAME = LATITUDE_quality_control
          PARAMETER QCFLAGVALS = 1,8
        An observation is considered valid if each of the specified flags takes
        a permitted value.
    - LOCATION_BASED_THINNING_TYPE
        XYZ* | XY | CELL | NIL
  Parameters specific to the reader:
    - ADDVAR (-)
        name of the variable to be added to the main variable (can be repeated)
    - SUBVAR (-)
        name of the variable to be subtracted from the main variable (can be
        repeated)
  Parameters common to all readers:
    - VARSCALE (-)
        scale factor (applied before VARSHIFT)
    - VARSHIFT (-)
        data offset to be added (e.g. -273.15 to convert from K to C)
    - FOOTRPINT (-)
        footprint of observations in km
    - MINDEPTH (-)
        minimal allowed model depth
    - MAXDEPTH (-)
        maximal allowed model depth
    - STRIDE (-)
        stride interval
    - YMIN (-)
        minimal allowed latitude
    - YMAX (-)
        maximal allowed latitude
    - ZMIN (-)
        minimal allowed depth
    - ZMAX (-)
        maximal allowed depth
\end{Verbatim}

Some details on the generic readers:
\begin{itemize}
\item \verb|TIMENAMES = <offset variable>, <base time variable>| is used when the observation time is the sum of a base time (with units of the form ``\verb|<units> since <date>|'') and offsets (in days, hours or seconds).
\item \verb|QCFLAGVALS| (values between 0 and 31) must immediately follow the corresponding \verb|QCFLAGNAME|.
\item The instrument tag is taken from (in order of precedence): parameter \verb|INSTRUMENT|; the global attribute named by \verb|INSTATTNAME|; attributes ``instrument'' (or ``Instrument'', ``INSTRUMENT'', ``platform'', etc.) of the data variable or global attributes with such names; the product name. The prefix \verb|INSTPREFIX|, if specified, is added to the tag found in the data file.
\item If the error variable (\verb|ESTDNAME|, by default \verb|error_std|) is absent, the observation error is taken from the attribute \verb|error_std| of the data variable, if present; otherwise it must be specified by \verb|ERROR_STD| in the observation data parameter file. If the dispersion variable (\verb|STDNAME|, by default \verb|std|) is present, the observation error is set to $\max(\sigma_{std}, \sigma_{error\_std})$.
\end{itemize}

The parameters listed above as ``common to generic readers'' differ somewhat between the generic readers (see the output of \verb|--describe-reader| for each reader):
\begin{itemize}
\item \verb|gridded_xy| reads 2D data and takes \verb|ZVALUE| (default 0) rather than \verb|ZNAME|; it also accepts \verb|NPOINTSNAME|, \verb|ADDVAR| and \verb|SUBVAR|;
\item \verb|gridded_xyz| takes \verb|ZNAME| rather than \verb|ZVALUE| and accepts \verb|NPOINTSNAME|;
\item \verb|gridded_xyh| requires \verb|GRIDNAME| (the name of the model grid the data is defined on, which is used instead of the longitude and latitude variables) and accepts \verb|NPOINTSNAME|;
\item \verb|profile| accepts \verb|EXCLUDEINST| (can be repeated);
\item \verb|ADDVAR| and \verb|SUBVAR| are only available in readers \verb|scattered| and \verb|gridded_xy|.
\end{itemize}
The custom readers have the following specific parameters: \verb|ADDBIAS| (\verb|amsr2|, \verb|navo|), \verb|MINWIND| and \verb|ORBITS| (\verb|amsre|), \verb|QCFLAGVALS| (\verb|cmems|).

\subsection{Superobing}
\label{sec:superobing}

``Superobing'' is the process of reduction of the number of observations by merging spatially close observations before their assimilation.
EnKF-C merges observations if:
\begin{itemize}
\item they belong to the same model grid cell (or superobing cell, see below) and the same layer;
\item are of the same type;
\item are from the same instrument (unless PREP is run with \verb|--superob-across-instruments|);
\item belong to the same batch (only if PREP is run with \verb|--no-superobing-across-batches|);
\item have the same footprint;
\item for asynchronous observations -- belong to the same time slot.
\end{itemize}
The horizontal size of superobing cells can be increased from the default of 1 model grid cell to $N \times N$ cells by setting \verb|SOBSTRIDE = <N>| in the main parameter file or (for a particular observation type) in the observation types parameter file; the vertical size is always equal to 1 layer.
\verb|SOBSTRIDE| must be $\ge 1$; superobing can be switched off by running PREP with \verb|--no-superobing|.

The observations are merged by averaging their values, coordinates and times with weights inversely proportional to the observation error variance.
The observation error variance of a superobservation is set to the inverse of the sum of inverse observation error variances of the merged observations.
The product and instrument fields of the superobservation are set either to those of the merged observations or to -1, depending on whether the merged observations have identical values for these fields or not.
(Because by default only observations from the same instrument are merged, the instrument of a superobservation can only be -1 with \verb|--superob-across-instruments|.)

Command line parameter \verb|--consider-subgrid-variability| switches on considering the subgrid variability by calculating standard deviation of the merged observations $\sigma\!_{sub}$ and using $\sigma\!_{obs} = \max(\sigma\!_{obs}, \sigma\!_{sub})$.
The calculation of $\sigma\!_{sub}$ is currently done in a rather crude way, assuming equal weights for all merged observations.

Note that during superobing EnKF-C thins observations with identical positions, assuming that those must been obtained from high-frequency instruments (e.g. moorings).
The thinning involves replacing the corresponding batch of observations by batch average values.
The type of thinning can be set by parameter \verb|LOCATION_BASED_THINNING_TYPE| in observation data parameter file and can be set to ``XYZ'' (thinning observations with coinciding X,Y,Z coordinates; default), ``XY'' (coinciding X,Y coordinates only), ``CELL'' (all observations within cell), or ``0'' (``no'', ``false''; no thinning).
The location based thinning can be switched off by the command line parameter \verb|--no-thinning|, or for observations of a particular data type only by adding flag \verb|PERMIT_LOCATION_BASED_THINNING = no| in the corresponding section in the observation types parameter file.

\section{CALC}

CALC is the second stage of data assimilation in EnKF-C.
It calculates 2D arrays of local ensemble mean update coefficients $\mb w$ and (for $m\subtxt{dyn} > 1$, i.e. for EnKF and Hybrid) ensemble anomalies transforms $\mb T$, which together define the local transforms $\mb X_5$ \eqref{x5}.

The name of the binary for CALC is \verb|enkf_calc|.
It has the following usage and options:
\begin{Verbatim}[frame=single,fontsize=\footnotesize]
>./bin/enkf_calc 
  Usage: enkf_calc <prm file> [<options>]
  Options:
  --allow-logspace-with-static-ens
      confirm that static ensemble is conditioned for using log space
  --describe-prm-format [main|model|grid|obstypes]
      describe format of a parameter file and exit
  --forecast-stats-only
      calculate and print forecast observation stats only
  --ignore-no-obs
      proceed even if there are no observations
  --point-logs-only
      skip calculating transforms for the whole grid and observation stats
  --print-memory-usage
      print memory usage by each process
  --single-observation <lon> <lat> <depth> <type> <inn> <std>
      assimilate single observation with these parameters
  --skip-bad-forecast-obs
      skip observations with invalid forecasts
  --use-existing-transforms
      skip calculating ensemble transforms; use existing transforms*.nc files
  --use-rmsd-for-obsstats
      use RMSD instead of MAD when printing observation stats
  --use-these-obs <obs file>
      assimilate observations from this file; the file format must be compatible
      with that of observations.nc produced by `enkf_prep'
  --version
      print version and exit
  --write-HE
      write ensemble observations to file "HE.nc"
\end{Verbatim}

The option \verb|--forecast-stats-only| can be used for quick calculation of the innovation statistics for a given background (or ensemble).
This can be used, for example, for obtaining the persistence statistics, that is, the innovation statistics for the previous analysis.

The option \verb|--single-observation| provides an easy way to conduct the so called single observation experiments.
Normally, this experiments would be conducted in the EnOI mode, calculating increment (option \verb|--output-increment| of \verb|enkf_update|) rather than analysis.
When run in the EnKF mode, the increment (or analysis, depending on specifications) for each member is calculated.
Note that in this mode CALC (over)writes \verb|observations.nc| with the single observation.
\verb|--single-observation| cannot be combined with \verb|--forecast-stats-only|, \verb|--use-these-obs| or \verb|--point-logs-only|; \verb|--point-logs-only| cannot be combined with \verb|--forecast-stats-only|.
CALC stops if the analysis time (\verb|DA_DAY|) recorded in \verb|observations.nc| differs from the time in the main parameter file; this check is skipped with \verb|--use-these-obs|.

Apart from the transforms, CALC writes the following files:
\begin{itemize}
\item it adds to \verb|observations.nc| variables \verb|Hx_f| and \verb|std_f| (forecast observation ensemble mean and spread), \verb|Hx_a| and \verb|std_a| (the same for the analysis), and, if the K-factor is used (Sec.~\ref{sec:kfactor}), replaces \verb|estd| by the moderated observation error, keeping the original values in \verb|estd_orig|; it also updates variable \verb|status|;
\item \verb|enkf_diag.nc| (or \verb|enkf_diag-<grid ID>.nc|) with the observation impact data (Sec.~\ref{sec:impact});
\item \verb|obsbatches.txt| with statistics on observation batches and (if \verb|BADBATCHES| is specified) \verb|badobsbatches.txt| with the detected bad batches (Sec.~\ref{sec:batches});
\item point logs (Sec.~\ref{sec:pointlogs});
\item with \verb|--write-HE|, file \verb|HE.nc| with the ensemble observations (variable \verb|HE(nmem, nobs)|); in EnOI and Hybrid modes it contains the combined ensemble, that is, with the static anomalies scaled according to \eqref{hybrid_combined} and added to the dynamic ensemble mean (global attribute \verb|nmem_dynamic| gives the number of dynamic members).
\end{itemize}
During calculations CALC uses a temporary directory \verb|.enkftmp| (with the default compile flags); if this directory exists at start (e.g. after a crashed run), CALC quits and the directory needs to be removed manually.
The same applies to UPDATE.

Note that the calculated transforms \emph{do not} incorporate inflation.
Inflation is applied during UPDATE according to specifications (Sec.~\ref{sec:capping}).

\subsection{Observation functions}
\label{sec:hfunctions}

Model estimations for observations of each type are calculated using observation functions specified for this type by entry \verb|HFUNCTION| in the observation types parameter file, e.g.:
\begin{Verbatim}[frame=single,fontsize=\footnotesize]
NAME = SLA
...
HFUNCTION = standard
...
\end{Verbatim}
The available functions for each observation type are specified by the variable \verb|allhentries| in \verb|calc/allhs.c|:
\begin{Verbatim}
typedef struct {
    int issurface;
    char* H_tag;
    H_fn H;
} H_entry;

H_entry allhentries[] = {
    {1, "standard", H_surf_standard},
    {1, "biased", H_surf_biased},
    {0, "standard", H_subsurf_standard},
    {0, "wsurfbias", H_subsurf_wsurfbias},
    {0, "lowmem", H_subsurf_lowmem},
    /*
     * (the corresponding observation type can be entered either as
     * "issurface = 1" or "issurface = 0")
     */
    {1, "vertsum", H_vertsum},
    {0, "vertsum", H_vertsum},
    {1, "vertwavg", H_vertwavg},
    {0, "vertwavg", H_vertwavg}
};
\end{Verbatim}

H-functions ``standard'' are used by default.
They perform 2D or 3D linear interpolation on the model grid associated with the observation type.

Function ``lowmem'' can be used instead of ``standard'' for observation types associated with 3D model variables in situations with very large models.
Unlike ``standard'' it does not read the whole 3D dump for a variable but proceeds by reading only two layers at a time.

Function ``biased'' targets 2D variables corrected for bias; it requires a second variable (the bias field) in the entry \verb|VAR| of the observation type.
Function ``wsurfbias'' was developed to offset the values of temperature in the mixed layer by the value of SST bias.
It requires a second (2D) variable in the entry \verb|VAR| and either \verb|MLD_THRESH| or \verb|MLD_VARNAME| (but not both); it works on structured grids only, and with \verb|MLD_THRESH| it works in the EnKF mode only.

Function ``vertsum'' is used to assimilate observations corresponding to the sum of a model variable over all layers (such as the total sea-ice concentration).
It can be used only for variables on sigma or hybrid grids.
Function ``vertwavg'' matches observations with a weighted average of a model variable over all model layers; the weights are given by a second variable in the entry \verb|VAR|.
(Currently only positive contributions to the weighted average are accounted for.)
Functions ``vertsum'', ``vertwavg'' and ``wsurfbias'' cannot be used for log-transformed variables (\verb|APPLYLOG|).

\subsection{Asynchronous DA / FGAT}
\label{sec:async}

An observation type can be specified as ``asynchronous'' by specifying entry \verb|ASYNC| in the observation types parameter file (Sec. \ref{sec:obstypesprm}), e.g.:
\begin{Verbatim}[frame=single,fontsize=\footnotesize]
NAME = SST
(...)
ASYNC = 1 centre time
(...)
\end{Verbatim}
The above means that SST observations are considered to be asynchronous with time bins of 1 day, and that the model output time is stored in variable \verb|time| of the model output files.
The second token (\verb|centre| or \verb|endpoint|, or simply \verb|c| or \verb|e|) specifies the alignment of the intervals relative to the analysis time and must be specified (there is no default); the third token specifies the name of the time variable in the model output files and is mandatory as well.
If, for example, the assimilation time is specified as ``6085.5 days since 1990-01-01'', then with \verb|centre| alignment the interval 0 is centred at the time of assimilation, i.e. will be from day 6085.0 to day 6086.0; interval -1 -- from day 6084.0 to day 6085.0, interval 1 -- from day 6086.0 to day 6087.0, and so on.
It is possible to shift the asynchronous intervals so that not the centre but the start of interval 0 is located at the time of assimilation.
In this case one needs to use qualifier ``endpoint'' after the length of the interval, i.e.
\begin{Verbatim}[frame=single,fontsize=\footnotesize]
NAME = SST
(...)
ASYNC = 1 endpoint time
(...)
\end{Verbatim}
The interval 0 will then be from day 6085.5 to day 6086.5.

The model dumps for each asynchronous interval are read from files with names \spverb|mem<xxx>_<alias>_<time shift>.nc| in the dynamic ensemble directory (for the EnKF and Hybrid) or \spverb|bg_<alias>_<time shift>.nc| in the background directory (for the EnOI), or from multi-record files (see below); here \verb|<alias>| is the entry \verb|ALIAS| of the observation type, which defaults to the (first) model variable name.
Here ``time shift'' is the interval ID (with the interval 0 being centred/starting at the analysis time).
If the corresponding model outputs for a dynamic member (or the background, in the case of EnOI) cannot be found, CALC stops with an error.
The members of static ensembles are always read synchronously (from \spverb|mem<xxx>_<alias>.nc| in the static ensemble directory, see Sec.~\ref{sec:hybrid-enkfc}).
This can be tracked from the CALC log file, e.g.:
\begin{Verbatim}[frame=single,fontsize=\footnotesize]
  calculating ensemble observations:
  2026-11-04 15:13:14
    allocating 5602960128 bytes for HE array (in SHMEM):
    allocating 5602960128 bytes for HE^T array (in SHMEM):
    distributing 192 iterations:
      52 processes get 4 or 3 iterations
    SLA |aaaa|aaaa|aaaa
    SST |aaaa|aaaa|aaaa|aaaa|aaaa|aaaa|aaaa|aaaa|aaaa|aaaa|aaaa|aaaa|aaaa|aaaa|aaaa|aaaa|aaaa|aaaa|aaaa|aaaa|aaaa|aaaa|aaaa|aaaa|aaaa|aaaa|aaaa|aaaa|aaaa|aaaa|aaaa|aaaa|aaaa|aaaa|aaaa|aaaa
    TEM |aaaa|aaaa|aaaa
    SAL |aaaa|aaaa|aaaa
    SIC |aaaa|aaaa|aaaa
    SLA2 |aaaa|aaaa|aaaa
    SLA3 |aaaa|aaaa|aaaa
\end{Verbatim}
The entries ``a'' mean that the observations are assimilated asynchronously and the corresponding files have been found.
These entries would be replaced by ``s'' if the observations were assimilated synchronously because of lacking the corresponding files.
(But in hybrid systems these members can be processed by workers and not manifested in the log printed by the master only.)
The vertical lines indicate the time slots for asynchronous DA; in the above example the DAW has 36 time slots for SST and 3 time slots for other observation types.
The entries ``.'' would indicate calculating synchronous ensemble observations.

\begin{figure}[h]
  \centering
  \includegraphics[width = \textwidth]{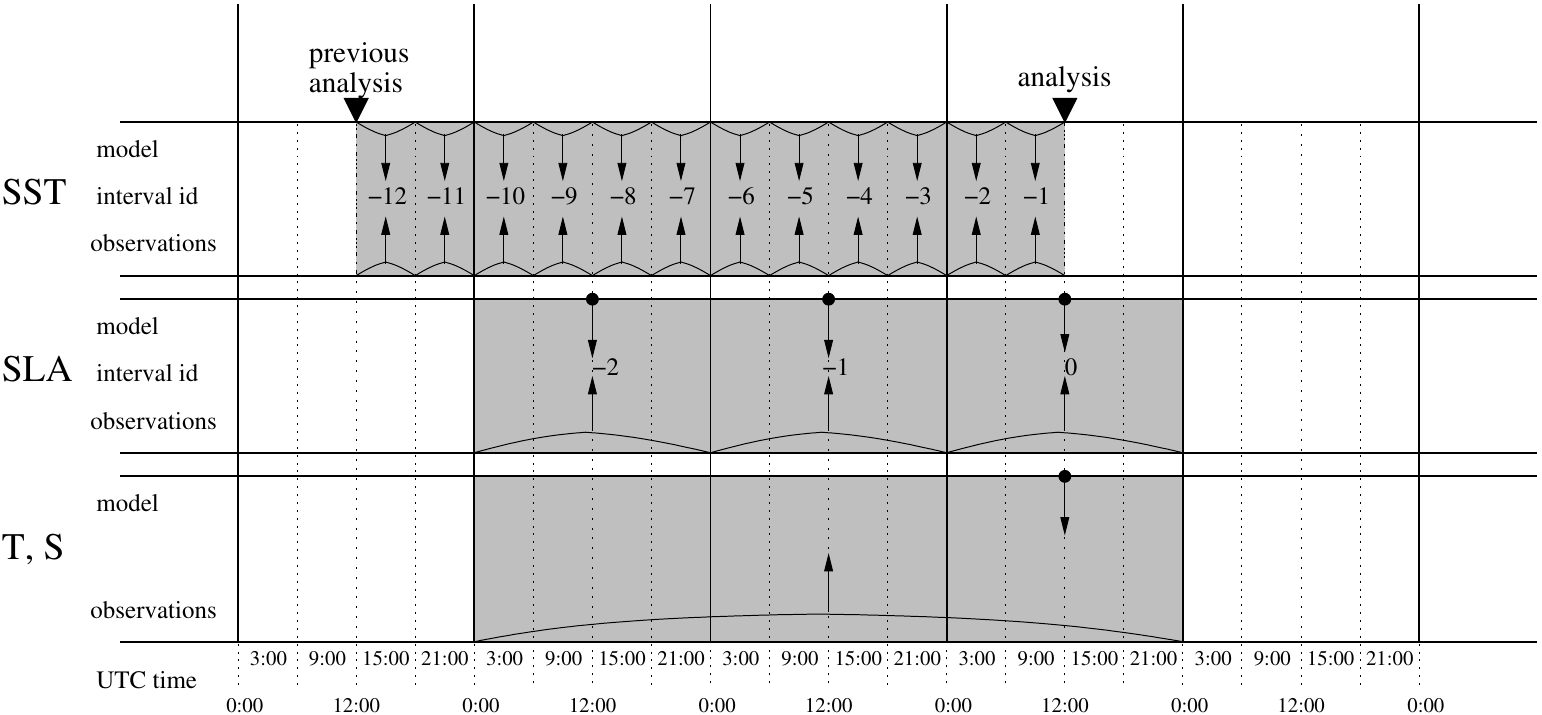}
  \caption{Example of observation timing in a MOM based ocean forecasting system.}
  \label{fig:mom_timing}
\end{figure}

Fig.~\ref{fig:mom_timing} shows an example of observation timing in a MOM based ocean forecasting system with a 3-day assimilation cycle.
In this system the ``fast'' SST data is assimilated asynchronously with 6\,h intervals using model fields averaged over these intervals; the slower SLA data is assimilated asynchronously with 24\,h intervals using daily model dumps; and in-situ T and S fields are assimilated synchronously.
This is achieved with the following settings in the observation types parameter file (any entry name starting with ``ASYNC'', e.g. \verb|ASYNCHRONOUS|, is accepted):
\begin{Verbatim}[frame=single,fontsize=\footnotesize]
NAME = SLA
VAR = eta_t
ISSURFACE = yes
ASYNCHRONOUS 1 C time
<...>
NAME = SST
VAR = temp sstb
ISSURFACE = yes
ASYNCHRONOUS 0.25 E time
<...>
NAME = TEM
VAR = temp sstb
ISSURFACE = no
<...>
NAME = SAL
VAR = salt
ISSURFACE = no
<...>
\end{Verbatim}

When an observation type is set as ``asynchronous'', EnKF-C aligns the centres of the time-binning intervals across the observation window with the model output times.

The second token after \verb|ASYNCHRONOUS| specifies the interval duration in days.
The third token specifies the alignment of these intervals relative to the analysis time, and the fourth token specifies the name of the time variable.

There are two alignment options the time binning intervals:
\begin{itemize}
\item[-] \verb|endpoint| (or \verb|E|): Aligns the analysis time with the start of the zeroth interval.
\item[-] \verb|centre| (or \verb|C|): Aligns the analysis time with the centre of the zeroth interval.
\end{itemize}

Model outputs can be associated with the time binning intervals either {\bf directly} or via a {\bf time search}:
\begin{enumerate}
\item {\bf Direct association:} The interval ID is embedded directly into the file name. The model output must be written using the template \spverb|printf("
The output time for interval $i$ must strictly satisfy $t_i = t_0 + i \, dt$, where $dt$ is the interval length and $t_0$ is the analysis time (for \verb|centre| alignment) or the analysis time plus $dt/2$ (for \verb|endpoint| alignment).
If a mismatch occurs between the interval and the model output time, CALC will exit with error.
\item {\bf Time search:} EnKF-C can also automatically locate the correct model output records by matching the interval times with the timestamps inside the files. To use this, write model outputs using the wildcard template \spverb|printf("
EnKF-C checks timestamps of all records matching this template and selects one with the correct time.
If no matching record is found, CALC exits with error.
\end{enumerate}
(For the EnOI background the templates are \spverb|"

\subsection{Interpolation of ensemble transforms}

Local ensemble transforms $\mb T$ and ensemble weights $\mb w$ (Sec. \ref{sec:enkf-overview}) represent smooth fields with the characteristic spatial variability scale of the localisation radius.
This makes it possible to reduce the computational load in CALC by calculating local transforms or weights on a subgrid with specified stride and then linearly interpolating the transforms or weights in the intermediate grid cells \citep{yan09a}.
The value of the stride is defined by the entry \verb|STRIDE| in the main parameter file and can be overwritten for a particular grid in the grid parameter file.

The transforms are written to \verb|transforms.nc| (\verb|transforms-<id>.nc| for multiple grids).
The file contains variable \verb|w(j, i, m)| for all modes and, if the dynamic ensemble size $m\subtxt{dyn} > 1$, variable \verb|T(j, i, m_dyn, m)|, where \verb|(j, i)| are the indices of the nodes of the strided grid; global attributes \verb|stride| and \verb|grid_name| specify the stride and the grid name.

\subsection{Adaptive moderation of observations}
\label{sec:kfactor}

One of the standard QC procedures in DA is the so called background check, when an observation is compared with the forecast and discarded if the innovation magnitude exceeds some threshold.
The downside of this approach is that it can not distinguish between situations of an outlier, big model error (e.g. because of an error in forcing), or model divergence.
While one probably would like to discard an outlier, it is usually desirable to make use of valid observations, although, perhaps, with a reduced impact, to avoid ``over-stressing'' the model.
In EnKF-C this is achieved by adaptive moderation of the observation impact by restricting the magnitude of the increment from a given observation in observation space by $K$ times magnitude of the forecast ensemble spread \citep{sak17a}.

Specifically, the adaptive moderation of the observation impact is conducted by smoothly increasing the observation error depending on the magnitude of innovation as follows:
\begin{align*}
  \sigma^2_{obs} \leftarrow \left[(\sigma_{f}^2 + \sigma_{obs}^2)^2 + \sigma_f^2 \, d^2 / K^2\right]^{1/2} - \sigma_f^2,
\end{align*}
where $\sigma_{obs}$ is the observation error standard deviation, $\sigma_f$ -- forecast ensemble spread, $d$ -- innovation, and $K$ -- the so called K-factor defined in the main parameter file (Sec. \ref{sec:mainprm}).
Tests with small models show that setting the $K \ge 2$ makes a marginal impact (if any) on performance of weakly suboptimal systems, while it still can be quite beneficial in situations with large innovations.

\subsection{Moderation of spread reduction}
\label{sec:rps}

The moderating parameter $\alpha \in (0, 1]$ specified in the main parameter file via the entry \verb|ALPHA| allows one to reduce the contraction of ensemble during assimilation, while leaving the increment unchanged (``relaxation to prior spread'', RPS, \citealt{zha04a}, eq.~5).
It modifies the right multiplied ensemble transform matrix as
\begin{align*}
  \mb T_R \leftarrow \mb I + \alpha (\mb T_R - \mb I).
\end{align*}
Setting $\alpha = 0$ results in no update of the ensemble anomalies, while $\alpha = 1$ results in full update.
RPS is only applied when the dynamic ensemble size is greater than 1.

\subsection{Innovation statistics}

In its course CALC calculates some basic innovation statistics: number of observations, mean absolute forecast innovation, mean absolute analysis innovation, mean forecast innovation, mean analysis innovation, mean forecast ensemble spread, and mean analysis ensemble spread.
This statistics is provided for each region defined in the main parameter file (Sec.~\ref{sec:mainprm}), as well as for each time slot defined for asynchronous DA, and for each instrument.
By default, EnKF-C defines one statistical region ``Global'' with extent $[x_1,x_2] = [-999,999], [y_1,y_2] = [-999,999]$.

In addition, for 3D observations CALC also calculates observation statistics in specified depth intervals.
These intervals can be set via the entry \verb|ZSTATINTS| in the main parameter file or (specific for a particular grid) in the grid parameter file.
By default, three intervals are defined: \verb|[0 DEPTH_SHALLOW]|, \verb|[DEPTH_SHALLOW DEPTH_DEEP]|, and \verb|[DEPTH_DEEP DEPTH_MAX]|, where \verb|DEPTH_SHALLOW| = 50, \verb|DEPTH_DEEP| = 500 and \verb|DEPTH_MAX| = 9999 are the macros defined in \verb|common/global.h|.

Following is an example of innovation statistics written to the log (standard output) of \verb|enkf_calc|:
\begin{Verbatim}[frame=single,fontsize=\footnotesize]
  printing observation statistics:
    region obs.type   # obs.  |for.inn.| |an.inn.|   for.inn.   an.inn.  for.spread  an.spread
    ------------------------------------------------------------------------------------------
    Tasman
           SLA          3003    0.067      0.038      0.033      0.012      0.035      0.025  
            -4           712    0.058      0.038      0.035      0.013      0.028      0.021  
            -3           785    0.093      0.040      0.060      0.019      0.052      0.034  
            -2           700    0.062      0.043      0.030      0.016      0.027      0.021  
            -1           668    0.049      0.031      0.017      0.004      0.028      0.021  
             0           138    0.078      0.033     -0.043     -0.016      0.045      0.029  
             j1         1323    0.070      0.033      0.041      0.016      0.037      0.024  
             n1          876    0.073      0.042      0.052      0.025      0.036      0.026  
             g1          785    0.054      0.042     -0.004     -0.009      0.029      0.024  
             N/A          19    0.101      0.037      0.097      0.031      0.059      0.036  
           SST          9316    0.346      0.174     -0.215     -0.094      0.358      0.254  
            -4          2946    0.327      0.166     -0.236     -0.092      0.342      0.245  
            -3          2733    0.368      0.183     -0.270     -0.133      0.362      0.256  
            -2          2560    0.352      0.169     -0.167     -0.057      0.370      0.262  
            -1           580    0.342      0.191     -0.148     -0.093      0.414      0.291  
             0           497    0.305      0.182     -0.126     -0.075      0.307      0.225  
             AVHRR      9316    0.346      0.174     -0.215     -0.094      0.358      0.254  
           TEM           768    0.581      0.365     -0.245     -0.151      0.320      0.251  
             ARGO        768    0.581      0.365     -0.245     -0.151      0.320      0.251  
             0-50m       125    0.418      0.230      0.049      0.027      0.365      0.281  
             50-500m     451    0.678      0.403     -0.266     -0.141      0.360      0.278  
             >500m       192    0.458      0.365     -0.387     -0.291      0.196      0.170  
           SAL           768    0.079      0.060      0.014      0.019      0.033      0.028  
             ARGO        768    0.079      0.060      0.014      0.019      0.033      0.028  
             0-50m       125    0.079      0.063      0.031      0.035      0.034      0.030  
             50-500m     451    0.092      0.067      0.026      0.032      0.039      0.032  
             >500m       192    0.048      0.041     -0.027     -0.021      0.018      0.016  
\end{Verbatim}
This excerpt shows innovation statistics for the region ``Tasman''.
It contains sections for SST, SLA and TEM observations.
The summary statistics for each observation type is shown at the top of each section; then statistics for days -4, -3, -2, -1 and 0 of a 5-day DAW are shown for the two asynchronous types, SST and SLA.
(More generally, the numbering of time intervals corresponds to their positions relative to the analysis time.
For more details see Section~\ref{sec:async}.)
After that, statistics for particular instruments is shown; ``N/A'' corresponds to superobservations resulted from merging observations from two or more instruments.
(From v1.115.0 there is no superobing across different instruments by default.)
For subsurface temperature also statistics for shallow (0--50\,m), deep ($>$500\,m), and intermediate (50--500\,m) observations is given.

The analysis innovation statistics is calculated from the updated (analysis) ensemble observations by CALC, thus avoiding the need to access analysis files produced later by UPDATE.
The update of ensemble observations is performed in the same way as that of any other element of the state vector:
for the EnKF -- by applying the appropriate local ensemble transforms to the forecast ensemble observations,
\begin{align*}
  \mathcal H (\mb E^a) \leftarrow \mathcal H (\mb E^f) \, \mb X_5;
\end{align*}
and for the EnOI -- by applying the appropriate local linear combination of the ensemble observation anomalies:
\begin{align*}
  \mathcal H (\mb E^a) \leftarrow \left[ \mathcal H (\mb x^f) + (\mb H \mb A^f) \mb w\right] \mb 1\T + \mb H \mb A^f.
\end{align*}

CALC can be used for calculating forecast observation statistics only (via command line option \verb|--forecast-stats-only|), without calculating transforms (EnKF) of update coefficients (EnOI).
In the EnKF or Hybrid modes this regime involves calculating the statistics for the ensemble observation spread (and therefore parsing of the forecast ensemble), while in the EnOI mode it only calculates the statistics for the forecast innovation (and therefore does not need to access the ensemble).

\subsection{Impact of observations}
\label{sec:impact}

In the course of its work CALC routinely calculates two metrics for assessing the impact of observations, degrees of freedom of signal (DFS) and spread reduction factor (SRF):
\begin{align*}
  & \mathrm{DFS} = \mathrm{tr}(\mb K \mb H) = \mathrm{tr}(\mb G \mb S), \\
  & \mathrm{SRF} = \sqrt{\frac{\mathrm{tr}(\mb H \mb P^f \mb H\T \mb R^{-1})}{\mathrm{tr}(\mb H \mb P^a \mb H\T \mb R^{-1})}} - 1 = \sqrt{\frac{\mathrm{tr}(\mb S\T \mb S)}{\mathrm{tr}(\mb G \mb S)}} - 1,
\end{align*}
where $\mathrm{tr}(\cdot)$ is the trace function.
The values of these metrics for each local analysis, calculated both for all observations and for observations of each type only, are written to files \verb|enkf_diag[-<grid ID>].nc| (variables \verb|dfs|, \verb|srf|, \verb|pdfs|, and \verb|psrf|; the files also contain the number of local observations \verb|nlobs| and its partial (per observation type) values \verb|pnlobs|).

Note that in EnKF-C DFS and SRF are calculated from the above expressions and represent theoretical values for the EnKF analysis; they coincide with the actual DFS and SRF values only for vanilla ETKF (i.e. ETKF with no inflation or relaxation to prior), but still are useful metrics for assessing the impact of observations in practice.

In the EnKF context DFS is a useful indicator of potential rank problems.
Normally, DFS calculated for assimilating observations over the characteristic time of the error growth should not exceed a fraction (a half, or better, a quarter) of the ensemble size.

SRF shows the ``strength'' of DA.
``Strong'' DA requires a nearly optimal system (otherwise it produces large artefacts and imbalances), which indeed never happens in practice.
Therefore, ideally, SRF should be small (below 1, on average).

\subsection{Multiple model grids}

EnKF-C permits using multiple model grids, in which case the ensemble transforms are calculated sequentially for each of the grids.
These transforms are then used for updating the model variables defined on the corresponding grids.

\subsection{Domains}
\label{sec:domains}

By default, all local observations for a given grid node contribute to the corresponding ensemble transform.
Sometimes it is desirable to disconnect observations of certain type from contributing to transforms on particular grids.
For example, it may be desirable in climate systems to disregard observations of the sea surface height (SSH) in updating the atmospheric variables.
The concept of ``domains'' introduced in v1.89.0 provides a mechanism for handling such situations within a single analysis.
It works as follows.
Each grid can be associated with a certain domain via the optional entry \verb|DOMAIN| in the grid parameter file.
For example, in a climate model one can have domains ``Ocean'' and ``Atmosphere''.
Then entry \verb|DOMAINS| in the observation types parameter file can list domains observations of this type are visible from.
By default, observations of any type are visible from all grids.

\subsection {``Multi-scale'' localisation}

It is possible to specify the localisation taper function as a linear combination of the Gaspari and Cohn's taper functions with different support radii:
\begin{align*}
  f(r) = \sum_{i=1}^N w_i f_0(\frac{r}{R_i}),
\end{align*}
where $w_i$ is the weight, $r$ is the distance, $R_i$ is the support radius, and
\begin{align*}
  f_0(\frac{x}{2}) = \left\{
  \begin{array}{ll}
    1 - \frac{5}{3} x^2 + \frac{5}{8} x^3 + \frac{1}{2} x^4 -\frac{1}{4} x^5, \quad & 0 \le x \le 1,\\
    -\frac{2}{3} x^{-1} + 4 - 5x + \frac{5}{3}x^2 + \frac{5}{8}x^3 - \frac{1}{2} x^4 + \frac{1}{12}x^5, \quad & 1 < x \le 2,\\
    0, \quad & 2 < x.
  \end{array}
  \right.
\end{align*}
This can be set by entries \verb|LOCRAD| and \verb|LOCWEIGHT| either in the main parameter file or in the observation types parameter file, e.g.:
\begin{Verbatim}
  LOCRAD 150 500
  LOCWEIGHT 0.9 0.1
\end{Verbatim}
(recall that entries in the observation types parameter file for particular observation types override the common settings in the main parameter file).
Note that the weights are normalised so that their sum is equal to 1.

\section{UPDATE}

UPDATE is the third and final stage of data assimilation in EnKF-C.
It updates the restart ensemble (EnKF, Hybrid) or the background (EnOI) by applying the transforms calculated by CALC.

The name of the binary for UPDATE is \verb|enkf_update|.
It has the following usage and options:
\begin{Verbatim}[frame=single,fontsize=\footnotesize]
>./bin/enkf_update
  Usage: enkf_update <prm file> [<options>]
  Options:
  --allow-logspace-with-static-ens
      confirm that static ensemble is conditioned for using log space
  --calculate-spread
      calculate forecast ensemble spread and write to spread.nc
  --describe-prm-format [main|model|grid]
      describe format of a parameter file and exit
  --direct-write
      write fields directly to the output file (default: write to tiles first)
  --no-fields-write
      do not write analysis fields (only diagnostic data)
  --no-update
      exclude tasks that require ensemble update
  --output-increment
      output analysis increment (default: output analysis)
  --write-inflation
      write capped inflation magnitudes to inflation.nc
  --version
      print version and exit
\end{Verbatim}

The analysed restarts are written to separate files using the same names as the forecast files but with an extra suffix \verb|.analysis| or \verb|.increment| (Sec.~\ref{sec:filenames}), depending on whether the analysis or analysis increment is written.

By default, UPDATE first writes each updated horizontal field of the model to a separate file (``tile'') in the temporary directory \verb|.enkftmp|, and then concatenates these fields into analysis files.
This approach is somewhat less effective than direct writing to analysis files (without intermediate tiles), but, unfortunately, the direct writing is generally not reliable due to parallel I/O issues with NetCDF.
Note that in some cases it proved to be possible to obtain robust performance with direct write using ``classic'' or ``64-bit-offset'' NetCDF formats.
Option \verb|--direct-write| with more than one process is refused for \verb|NCFORMAT = NETCDF4| (the default).
When run on a single process UPDATE uses direct write by default, unless \verb|FIELDSPLIT| $> 1$.
If the directory \verb|.enkftmp| exists at start, UPDATE quits.

Notes on other options:
\begin{itemize}
\item \verb|--write-inflation| is silently cancelled when there is no dynamic ensemble (EnOI) or the inflation is \verb|PLAIN| in the main parameter file.
\item \verb|--no-update| also cancels writing of the inflation and of the analysis to point logs.
\end{itemize}

Updating a horizontal field requires reading the full ensemble into memory.
This may require large available memory, particularly with very large models.
From version 2.43.0 it is possible to calculate analyses on fractions of a horizontal field, which may provide better scalability.
This option is specified by parameter FIELDSPLIT of the main parameter file.
When FIELDSPLIT is set to a number greater than one, an additional stage of combining outputs on these fractional tiles into full fields is conducted.
\verb|FIELDSPLIT| $> 1$ is incompatible with \verb|FIELDBUFFERSIZE| $> 1$ and with direct write; it is not supported by \verb|ens_diag|.

\subsection{Capping of inflation}
\label{sec:capping}

Applying spatially uniform ensemble inflation involves areas with no local observations, where no assimilation is conducted.
It can gradually inject energy into the model and deteriorate performance of the DAS over time.
Similar problems may arise due to lack of correlation between some state elements updated with the same transforms, so that even in presence of local observations the ensemble spread for some elements may hardly reduce after assimilation, yet the ensemble anomalies are inflated.

To avoid this behaviour EnKF-C currently restricts inflation by specified fraction (1 by default) of the the spread reduction factor calculated directly for each element of the state vector during the update.
For example, if inflation is specified as
\begin{Verbatim}
INFLATION = 1.06 0.5
\end{Verbatim}
then the ensemble anomalies for any model state element will be inflated by 6\,\%, but no more than $1 + 0.5 (\sigma_f / \sigma_a - 1)$, where $\sigma_f$ and $\sigma_a$ represent the forecast and analysis ensemble spreads for this element.
Specification
\begin{Verbatim}
INFLATION = 1.06
\end{Verbatim}
is equivalent to
\begin{Verbatim}
INFLATION = 1.06 1
\end{Verbatim}
Capping inflation by the magnitude of reduction of the ensemble spread is the default in EnKF-C; to revert to the uniform inflation add qualifier \verb|PLAIN| to the entry \verb|INFLATION| in the main parameter file, e.g.:
\begin{Verbatim}
INFLATION = 1.06 PLAIN
\end{Verbatim}
The common inflation settings in the main parameter file can be overwritten by settings for particular model variables specified in the model parameter file (Sec.~\ref{sec:modelprm}), using the same syntax.

\section{Hybrid covariance}
\label{sec:hybrid-enkfc}

From v2.0.0 EnKF-C makes it possible using hybrid state error covariance by combining covariances from the EnKF ensemble and an ensemble of static anomalies (Sec.~\ref{sec:hybrid-theory}); from v2.49.0 several static ensembles can be used.
This option is activated by specifying \verb|MODE = HYBRID| in the main parameter file, the directory of each static ensemble, and, for each static ensemble, the mixing coefficient \verb|GAMMA| (default 1; see Sec.~\ref{sec:mainprm}).
The static anomalies are calculated by subtracting the mean of each static ensemble; therefore static ensembles do not need to be centred.

When the mode is set to ``hybrid'', the ensemble spread written in the innovation statistics summary at the end of CALC is that of the combined ensemble \eqref{hybrid_combined}, normalised by $m - N - 1$ \eqref{cov_combined}.
Similarly, the ensemble spread fields calculated during UPDATE by specifying option \verb|--calculate-spread| is the ensemble spread of the combined ensemble, with the same normalisation (for the EnKF this means normalisation by $m - 1$).
To calculate spread of dynamic ensemble only one needs to set \verb|MODE = EnKF|.
To calculate innovation statistics with forecast ensemble spread of the dynamic ensemble only one needs to set \verb|MODE = EnKF| and rerun CALC with option ``--forecast-stats-only''.

Note that \verb|GAMMA| must be positive (\verb|GAMMA = 0| is rejected; to switch off the static covariance use \verb|MODE = EnKF|), while setting \verb|ENSSIZE_DYNAMIC = 1| makes the hybrid system formally equivalent to the EnOI (each static ensemble must have at least 2 members).
A static member value that is not valid at some grid point is replaced by the forecast mean (that is, this member has zero anomaly at this point).

\subsection{On the asynchronous DA in a hybrid system}

In a hybrid system the ensemble observation anomalies for the static part of the ensemble are calculated from the static ensemble, i.e. without considering observation time.
To form the full ensemble observations they are then added to the corresponding ensemble mean observations from the dynamic ensemble.
If the ensemble observations of the dynamic ensemble are asynchronous then the ensemble mean is also asynchronous.
Therefore, the static part of the ensemble functions similarly to an FGAT EnOI system, while the dynamic part functions as an asynchronous EnKF.

\section{Ensemble diagnostics}
\label{sec:diag}

Starting from version 2.27.0 a new binary \verb|enkf_diag| was added to EnKF-C; from v.2.37.0 it is called \verb|ens_diag|.
It aims at calculating various ensemble diagnostics outside the context of DA cycling.
It has the following usage and options:
\begin{Verbatim}[frame=single,fontsize=\footnotesize]
>./bin/ens_diag
  Usage: ens_diag <prm file> [<options>]
  Options:
  --calculate-spread
      calculate ensemble spread and write to spread.nc
  --calculate-vertical-correlations
      calculate correlation coefficients between surface and other layers of
      3D variables and write to vcorr.nc
  --calculate-vertical-correlations-with <varname1> <layer1>
    [<varname2> <layer2>] [...]
      calculate correlation coefficients between specified field (a layer of
      a variable) and all other fields on the same horizontal grid and
      write to vcorr-<varname>-<layer>.nc
  --calculate-vertical-covariances-with <varname1> <layer1>
    [<varname2> <layer2>] [...]
      calculate covariances between specified field and all other fields
      on the same horizontal grid and write to vcov-<varname>-<layer>.nc
  --calculate-vertical-sensitivities-with <varname1> <layer1>
    [<varname2> <layer2>] [...]
      calculate sensitivities between specified field and all other fields
      on the same horizontal grid and write to vsens-<varname>-<layer>.nc
  --describe-prm-format [main|model|grid]
      describe format of a parameter file and exit
  --version
      print version and exit
\end{Verbatim}
While most of the options above are straightforward, we note that the spread and the covariances are normalised by $m - N - 1$ (Sec.~\ref{sec:hybrid-theory}), with static ensembles (in EnOI and Hybrid modes) treated in the same way as in CALC and UPDATE; and that the option ``--calculate-vertical-sensitivities-with'' calculates the ratio $s_{xy} = \sigma_{x,y} / \sigma_{y,y}$, where $\sigma_{x,y}$ is the covariance between $x$ and $y$, and $\sigma_{y,y}$ is the variance of $y$, so that the increment of $x$ is related to the increment of $y$ as $\Delta x = s_{xy} \Delta y$ \citep[see e.g.][eq.\,5]{and03a}.

Note that \verb|ens_diag| can work with a much simpler parameter file than
other executables in EnKF-C.
It requires \verb|FIELDSPLIT = 1|.
Because \verb|ens_diag| does not have option \verb|--allow-logspace-with-static-ens|, it currently cannot be run for model variables with \verb|APPLYLOG| in EnOI or Hybrid modes.

\section{DA tuning}
\label{sec:datuning}

Following are the main parameters for DAS tuning in EnKF-C:
\begin{itemize}
\item R-factors (Sec.~\ref{sec:obstypesprm});
\item localisation radii (Sec.~\ref{sec:obstypesprm});
\item inflation magnitudes (Sec.~\ref{sec:modelprm},\ref{sec:capping});
\item K-factor (Sec.~\ref{sec:kfactor});
\item magnitude of relaxation to prior spread (parameter \verb|ALPHA|, Sec.~\ref{sec:rps});
  \item scaling of static covariance (parameter \verb|GAMMA|, Sec.~\ref{sec:hybrid-enkfc}).
\end{itemize}

The R-factors can be defined for each observation type.
They represent scaling coefficients for the corresponding observation error variances and affect the impact of these observations: increasing R-factor decreases the impact of observations and vice versa.
Specifying R-factor equal $k$ produces the same increment as reducing the ensemble spread by $k^{1/2}$ times.

The main parameter file defines the base R-factor common for all observation types.
It is possible to specify additional R-factors for observations of each type (Sec. \ref{sec:obstypesprm}); the resulting R-factor for an observation is then given by multiplication of the common R-factor and the additional R-factor specified for observations of this type.

Multiplicative inflation can be seen as an additional forgetting factor in the KF.
In EnKF-C one can specify the inflation multiple for analysed ensemble anomalies, e.g.:
\begin{Verbatim}[frame=single,fontsize=\footnotesize]
> grep INFLATION main.prm
INFLATION = 1.05 PLAIN
> grep temp -A 1 model.prm
VAR = temp
INFLATION = 1.07 PLAIN
\end{Verbatim}
In this case all model variables except ``temp'' will have inflation of 5\,\%, while ``temp'' will have inflation of 7\,\%.
The ability to define different inflation rates for different variables can be useful for non-dynamical variables, such as estimated biases, helping to avoid the ensemble collapse for them.
In general, to retain dynamical balances one should rather avoid using different inflation magnitudes across model variables.
Note that even small inflation can substantially affect the ensemble spread established in the course of evolution of the system.
By default EnKF-C applies adaptive capping of inflation (Sec.~\ref{sec:capping}).

Localisation radius is defined by the entry \verb|LOCRAD| in the parameter file.
Specifically, this entry defines the support localisation radius (in km).
This is different to the ``effective'' localisation radius, which is defined sometimes as $e^{1/2}\approx 1.65$ - folding distance.
For the Gaspari and Cohn's taper function used in EnKF-C the effective radius is approximately 3.5 times smaller than the support radius.

Increasing the localisation radius increases the number of local observations and hence the overall impact of observations.
To compensate this in a system with horizontal localisation one has to change the R-factor as the square of the localisation radius.

From v1.77.0 it is possible to limit the maximal number of local observations of each observation type via the entry \verb|NLOBSMAX| in the main parameter file (Sec.~\ref{sec:mainprm}).
This common setting can be overriden for particular observation types in the observation types parameter file (Sec.~\ref{sec:obstypesprm}).
Note that using this setting can result in discontinuity of the analysis because the set of observations used for local analyses in adjacent grid cells can change in a discontinuous way.
Also, it forces sorting of the local observations by distance in CALC, which can substantially increase the search time.
Therefore the general advise is to avoid using this functionality except perhaps interpolation oriented products.

The adaptive moderation of observation impact with the K-factor (Sec. \ref{sec:kfactor}) is normally fairly non-intrusive, while can be essential for robust performance in situations with large innovations.
We suggest setting \verb|KFACTOR = 2| as default, and reducing it to 1 if the analysis turns out not balanced enough.

\section{Point logs}
\label{sec:pointlogs}

It is often desirable to investigate the drivers of the analysis or, more generally, certain features of the DAS and their behaviour over time.
In practice such investigations can be logistically complicated due to limitations on storage and/or access to it.
Yet, it is usually feasible to save the model state and observations for a number of specified locations.

EnKF-C provides capability of saving complete DA related information for specified horizontal locations in so called ``point logs''.
The locations are specified in the main parameter file, e.g.:
\begin{Verbatim}
POINTLOG 154.3 -34.1
POINTLOG 158.39 -31.7
\end{Verbatim}
Here the information will be saved for points with geographic coordinates (154.3,-34.1), (158.39, -31.7) in files \verb|pointlog_154.300,-34.100.nc|, \verb|pointlog_158.390,-31.700.nc|, and so on.
A point log location must be inside at least one of the model grids; otherwise CALC stops with an error.
By default the ensemble transforms and all forecast and analysis state variables existing at these locations are saved; however, if an optional grid name is specified as the third parameter of the \verb|POINTLOG| entry, e.g.
\begin{Verbatim}
POINTLOG 154.3 -34.1 t-grid
\end{Verbatim}
then only transforms and variables for this grid are saved.

Following is an example of point log file header (file \verb|pointlog_156.000,-32.000.nc| from example~4 after running \verb|make hybrid|).
In the EnKF mode (no static ensembles) a single ensemble dimension \verb|m| is used instead of \verb|m1| and \verb|m2|.
Attributes \verb|SCHEME| and \verb|ALPHA| are only written when the dynamic ensemble size is greater than 1; with more than one static ensemble the attribute \verb|GAMMA| is replaced by attributes \verb|ENSSIZE_STATIC<s>| and \verb|GAMMA<s>| for each static ensemble \verb|s| (counted from 0).
For a grid defined with \verb|HGRIDFROM| the grid variable also has attribute \verb|hid| (the ID of the grid whose horizontal grid is used).
\begin{Verbatim}[frame=single,fontsize=\footnotesize]
netcdf pointlog_156.000\,-32.000 {
dimensions:
	m1 = 96 ;
	m2 = 240 ;
	p = 1440 ;
	p0 = 1440 ;
	p1 = 1440 ;
	nk0 = 2 ;
	nk1 = 2 ;
variables:
	char dimension_reference ;
		dimension_reference:m1 = "size of dynamic ensemble" ;
		dimension_reference:m2 = "total size of ensemble" ;
		dimension_reference:p = "total number of observations" ;
		dimension_reference:p0 = "number of observations on grid 0" ;
		dimension_reference:p1 = "number of observations on grid 1" ;
		dimension_reference:nk0 = "number of layers in grid 0" ;
		dimension_reference:nk1 = "number of layers in grid 1" ;
	int obs_ids(p) ;
		obs_ids:long_name = "observation ID in observations.nc" ;
	float obs_lcoeffs(p) ;
		obs_lcoeffs:long_name = "observation tapering coefficient" ;
	float obs_lon(p) ;
		obs_lon:long_name = "observation longitude" ;
	float obs_lat(p) ;
		obs_lat:long_name = "observation latitude" ;
	float obs_depth(p) ;
		obs_depth:long_name = "observation depth" ;
	float obs_val(p) ;
		obs_val:long_name = "observation value" ;
	float obs_estd(p) ;
	float obs_fi(p) ;
		obs_fi:long_name = "observation fractional grid index i" ;
	float obs_fj(p) ;
		obs_fj:long_name = "observation fractional grid index j" ;
	float obs_fk(p) ;
		obs_fk:long_name = "observation fractional grid index k" ;
	int obs_type(p) ;
		obs_type:long_name = "observation type ID" ;
		obs_type:SLA = 0 ;
		obs_type:SST = 1 ;
		obs_type:TEM = 2 ;
		obs_type:SAL = 3 ;
		obs_type:RFACTOR_SLA = 2. ;
		obs_type:LOCRAD_SLA = 200. ;
		obs_type:WEIGHT_SLA = 1. ;
		obs_type:GRIDID_SLA = 0 ;
		obs_type:RFACTOR_SST = 32. ;
		obs_type:LOCRAD_SST = 200. ;
		obs_type:WEIGHT_SST = 1. ;
		obs_type:GRIDID_SST = 0 ;
		obs_type:RFACTOR_TEM = 8. ;
		obs_type:LOCRAD_TEM = 800. ;
		obs_type:WEIGHT_TEM = 1. ;
		obs_type:GRIDID_TEM = 0 ;
		obs_type:RFACTOR_SAL = 8. ;
		obs_type:LOCRAD_SAL = 800. ;
		obs_type:WEIGHT_SAL = 1. ;
		obs_type:GRIDID_SAL = 0 ;
	int obs_inst(p) ;
		obs_inst:long_name = "observation instrument ID" ;
		obs_inst:j1 = 0 ;
		obs_inst:n1 = 1 ;
		obs_inst:ESACCI = 2 ;
		obs_inst:WindSat = 3 ;
		obs_inst:WMO851 = 4 ;
		obs_inst:WMO846 = 5 ;
	float obs_time(p) ;
		obs_time:long_name = "observation time" ;
		obs_time:units = "days from 6565.5 days since 1990-01-01" ;
	int grid-0 ;
		grid-0:id = 0 ;
		grid-0:name = "t-grid" ;
		grid-0:domain = "ALL" ;
		grid-0:fi = 49.5 ;
		grid-0:fj = 49.5 ;
		grid-0:nk = 2 ;
		grid-0:model_depth = 4642.25f ;
	int grid-1 ;
		grid-1:id = 1 ;
		grid-1:name = "c-grid" ;
		grid-1:domain = "ALL" ;
		grid-1:fi = 48.9999691741445 ;
		grid-1:fj = 49.0000077064579 ;
		grid-1:nk = 2 ;
		grid-1:model_depth = NaNf ;
	float s0(p0) ;
		s0:long_name = "standardised innovation on grid 0" ;
	float S0(m2, p0) ;
		S0:long_name = "standardised ensemble observation anomalies on grid 0" ;
	double w0(m2) ;
		w0:long_name = "increment coefficients on grid 0" ;
	double T0(m1, m2) ;
		T0:long_name = "ensemble anomalies transform on grid 0" ;
	float s1(p1) ;
		s1:long_name = "standardised innovation on grid 1" ;
	float S1(m2, p1) ;
		S1:long_name = "standardised ensemble observation anomalies on grid 1" ;
	double w1(m2) ;
		w1:long_name = "increment coefficients on grid 1" ;
	double T1(m1, m2) ;
		T1:long_name = "ensemble anomalies transform on grid 1" ;
	float eta_t(m2) ;
		eta_t:gridid = 0 ;
	float eta_t_an(m1) ;
		eta_t_an:gridid = 0 ;
		eta_t_an:INFLATION = 1.1f, 1.f ;
	float temp(nk0, m2) ;
		temp:gridid = 0 ;
	float temp_an(nk0, m1) ;
		temp_an:gridid = 0 ;
		temp_an:INFLATION = 1.1f, 1.f ;
	float salt(nk0, m2) ;
		salt:gridid = 0 ;
	float salt_an(nk0, m1) ;
		salt_an:gridid = 0 ;
		salt_an:INFLATION = 1.1f, 1.f ;
	float u(nk1, m2) ;
		u:gridid = 1 ;
	float u_an(nk1, m1) ;
		u_an:gridid = 1 ;
		u_an:INFLATION = 1.1f, 1.f ;
	float v(nk1, m2) ;
		v:gridid = 1 ;
	float v_an(nk1, m1) ;
		v_an:gridid = 1 ;
		v_an:INFLATION = 1.1f, 1.f ;

// global attributes:
		:lon = 156. ;
		:lat = -32. ;
		:SCHEME = "DEnKF" ;
		:ALPHA = 1. ;
		:GAMMA = 0.25 ;
		:ngrids = 2 ;
		:EnKF-C\ version = "2.50.0" ;
n                :analysis\ time = "6565.5 days since 1990-01-01" ;
                :wdir = "/home/599/pxs599/src/enkf-c/enkf/examples/4" ;
                :CALC\ command = "./enkf_calc hybrid.prm" ;
                :CALC\ start\ time = "2026-11-14 20:13:28" ;
                :output = "analysis" ;
                :UPDATE\ command = "./enkf_update --calculate-spread --write-inflation hybrid.prm" ;
                :UPDATE\ start\ time = "2026-11-14 20:14:02" ;
}
\end{Verbatim}

This data makes it possible to check DA algorithms by reproducing the ensemble transforms (for EnKF) or weights (for EnOI) calculated by EnKF-C from $\mb S$ and $\mb s$ according to Section~\ref{sec:numerical}; restore observations from $\mb s$ by using the corresponding R-factors and localisation coefficients; to monitor the ensemble spread for each model variable; calculate inflation applied to the analysed anomalies; calculate impacts of particular observations; and so on.

Note that in a multi-domain setting (Sec.~\ref{sec:domains}) the number of local observations seen on a particular grid (e.g. \verb|p0|) can be smaller than the total number of local observations $p$.
In this case to get the local observations on this grid one needs to filter out observations of types defined on domains other than the domain the grid belongs to.

\section{Use of innovation statistics for model validation}

EnKF-C can calculate innovation statistics for validating a model against observations only, without data assimilation.
The prerequisites are (i) observations and (ii) model dump readable by the code, and possibly (iii) auxiliary files for projecting the model state to observation space (e.g. grid specs and mean SSH).
To get the innovation statistics one needs to:
\begin{itemize}
\item set up the parameter files in a normal way (\verb|MODE = EnOI|), omitting the ensemble directory and assimilation related parameters;
\item run \verb|enkf_prep|;
\item run \verb|enkf_calc| with additional parameter \verb|--forecast-stats-only|.
\end{itemize}
The results will be written to the log of \verb|enkf_calc|. 
An example of using this functionality is available by running \verb|make stats| in \verb|examples/1| (see Sec.~\ref{example1}).
Note that on some machines parallel processing of multiple cycles may require compiling CALC with no \verb|-DMPI| flag (\verb|MPISTATUS_CALC =| in the Makefile); otherwise the jobs may be assigned to the same CPUs.

\section{Bias correction}
\label{sec:bias}

It is possible to estimate and correct bias for a model variable with the EnKF by generating and using an ensemble of bias fields.
These bias fields need to be subtracted from the corresponding observation forecasts.
This is accomplished by specifying a secondary variable in the observation type entry \verb|VAR| and by passing the name of this variable to the corresponding observation (H-) functions, which need to take care for subtracting the bias from the model forecasts.
As of v1.98.0, there are two such H-functions: \verb|H_surf_biased()| identified by entry ``biased'' in observation types parameter file, and \verb|H_subsurf_wsurfbias()|, identified by entry ``wsurfbias''.
The latter applies surface bias field to the mixed layer, which is defined as the layer where the variable involved deviates within specified threshold from the surface value.
(In the case of SST bias the common value for the MLD threshold is 0.2\,K.)

Because bias fields are usually assumed to persist (not change) during propagation, one may need to make specific settings for their inflation to avoid their collapse (loss of spread) over time.
Another possibility is to introduce a ``forgetting'' stochastic model for bias fields, for example:
\begin{align*}
  \mb x_{i+1} = \lambda \, \mb x_i + (1 - \lambda^2)^{1/2} \ms \sigma,
\end{align*}
where $\lambda$ is the forgetting factor, $0 \le \lambda \le 1, \; 1 - \lambda \ll 1$, and $\ms \sigma \sim \sigma_0 N(0, 1)$, where $\sigma_0$ is the error standard deviation of $\mb x$.
This can be specified for a model variable via entry \verb|RANDOMISE <lambda> <sigma_0>| in the corresponding section of the model parameter file (Sec.~\ref{sec:modelprm}).
It is applied by UPDATE to the analysed dynamic ensemble members (before the increment is formed, so that it is also reflected in the output with \verb|--output-increment|).
In the current implementation the random perturbation is a single (field-wide) value for each variable and dynamic member, the same for all grid points and layers; the mean of the perturbations over the dynamic ensemble is removed; and masked (land) points are not perturbed.
The random numbers are generated on the master process (see Sec.~\ref{sec:seed} for reproducibility).

\section{Assimilation in log space}
\label{sec:logspace}

The entry \verb|APPLYLOG| in the model parameter file makes it possible to conduct assimilation for a variable in log space.
This means that a transform with log10 function will be applied to model values and observations before DA, and the inverse transform with pow10 will be applied after DA.
The observations are log-transformed by PREP; observations with non-positive values are discarded.
This option can be applied only for positive variables.
For observation types associated with log-transformed variables the observation error must be specified explicitly in the observation data parameter file (entry \verb|ERROR_STD|, which is then interpreted in log space), and \verb|OFFSET| cannot be used.

To apply log10/pow10 transforms in EnOI or Hybrid modes (\verb|MODE = ENOI| or \verb|MODE = HYBRID|) is only possible if the static ensemble is in logarithmic space.
This must be confirmed by the user by using option \verb|--allow-logspace-with-static-ens|.

When APPLYLOG is specified, the ensemble spread and ensemble vertical correlations are calculated for the transformed variable.

\section{System issues}

\subsection{Compiler flags}
\label{sec:flags}

Following is a brief description of the compiler flags in EnKF-C.
The flags are set in the \verb|Makefile| (variables \verb|PREPCALC_FLAGS|, \verb|CALC_FLAGS|, \verb|UPDATE_FLAGS| and \verb|DIAG_FLAGS|); the flags in effect for a binary are printed by running it with option \verb|--version|.

\begin{description}
\item{\verb|MPI|} (CALC, UPDATE, ens\_diag)
  Builds the parallel (MPI) version of a program.
  It is set via variables \verb|MPISTATUS_CALC|, \verb|MPISTATUS_UPDATE| and \verb|MPISTATUS_DIAG| in the Makefile (default: set); PREP is serial only.
\item{\verb|INTERNAL_QSORT_R|} (PREP, CALC)
  Uses internal (portable) code for \verb|qsort_r()|.
  Has to be defined for compiling on platforms without GNU \verb|qsort_r()| (e.g. Mac OS).
  (In versions before 2.50.0 it only took effect for PREP.)
\item{\verb|SHUFFLE_ROWS|} (CALC, MPI only)
  Supposed to produce more latitudinally uniform load between CPUs.
  Currently, because transforms for each row of the grid are sent to the master process for writing, this option effectively makes no difference to performance, I believe.
  It is ignored when \verb|TW_VIAFILE| or \verb|USE_MPIQUEUE| is defined.
\item{\verb|USE_SHMEM|} (CALC, MPI only)
  Uses shared memory for storing ensemble observations.
  This functionality requires MPI-3.
  It reduces the memory footprint by storing one instance of large objects per compute node.
  From v1.110.0 EnKF-C also uses shared memory for storing grid K-D trees and observation K-D trees, and from v1.111.11 -- for storing the observation array.
  This is a default option, but can be unset, particularly for smaller systems, when memory is not an issue.
  This option requires that the MPI ranks on each compute node are contiguous (``block'' placement, e.g. \verb|mpirun --map-by core| or \verb|--map-by slot|, which is the default in most MPI implementations); CALC checks this at start and quits otherwise (e.g. with \verb|--map-by node|).
\item{\verb|MINIMISE_ALLOC|} (CALC)
  Pre-allocates arrays in CALC to reduce potential problems with memory fragmentation.
  This is a default option from v1.103.0.
\item{\verb|OBS_SHUFFLE|} (CALC)
  Randomly shuffles observations before parsing them into K-D trees.
  Potentially this can substantially improve performance in the case of spatially ordered observations (not verified).
\item{\verb|TW_VIAFILE|} (CALC, MPI only)
  Communicate ensemble transforms via files (in directory \verb|.enkftmp|).
  Use this option if MPI communication becomes clogged.
  It is defined automatically when CALC is compiled with \verb|USE_MPIQUEUE|.
\item{\verb|DEFLATE_ALL|} (CALC, UPDATE)
  Apply specified NetCDF deflation to all NetCDF files, including ensemble transforms, spread, inflation, and various tiles.
  This can save some disk space, but slows down i/o, particularly assembling.
\item{\verb|USE_MPIQUEUE|} (CALC, UPDATE, MPI only)
  Distributes jobs in the main cycle to workers on ``as it goes'' basis rather than by assigning predefined number of iterations.
  This is particularly useful if the jobs are unbalanced.
  (In CALC a single job is calculating transforms for a row of a horizontal grid; in UPDATE it is applying transforms to the ensemble of horizontal fields at a specific level.)
  This is a default option for both CALC and UPDATE; consider removing it when running on just a few CPUs.
  In CALC it implies \verb|TW_VIAFILE| and cancels \verb|SHUFFLE_ROWS|.
\item{\verb|NCW_SKIPSINGLE|} (UPDATE, ens\_diag)
  Skips ``normal'' (not unlimited) ``inner'' dimensions of length one when copying definitions of variables from one NetCDF file to another.
\end{description}

\subsection{Memory footprint}
\label{sec:memory}

To reduce the memory footprint, most of the potentially large arrays in EnKF-C use \verb|float| data type.

The following table lists the most memory-wise important objects.

{
  \renewcommand{\arraystretch}{1.2}
  \begin{tabular}{|l|c|c|c|c|c|}
    \hline
    Object & Typical size$^1$ & Typical size$^2$ & PREP & CALC & UPDATE\\
    \hline
    observation array & 6\,GB & 8\,GB & \textbullet & &\\
    super-observation array$^{2}$ & 1.5\,GB & 3\,GB & \textbullet & \textbullet &\\
    ensemble observations$^3$ $\mathcal H(\mb E)$ & 5\,GB $\times$ 2 & 21\,GB $\times$ 2 & & \textbullet &\\
    single grid$^3$ & 0.4\,GB$^4$ ($\times$ 2)& 1.6\,GB ($\times$ 3) & \textbullet & \textbullet &\\
    observation K-D trees$^3$ & 0.85\,GB & 2\,GB & & \textbullet &\\
    one 3D model field & 1\,GB & 4.6\,GB & & \textbullet &\\
    ensemble of one horizontal field & 2\,GB & 10\,GB & & & \textbullet\\
    transform array (file) & 22\,GB$^5$ ($\times$ 2)& 29\,GB$^5$ ($\times$ 3) & & & \textbullet\\
    \hline
  \end{tabular}\\[1mm]
  {\scriptsize$^{(1)}$\,for EnKF/OFAM3 system ($3600 \times 1500 \times 51$ grid, 96 members, $5\cdot 10^7$ observations, $1.3 \cdot 10^7$ super-observations)}\\
  {\scriptsize$^{(2)}$\,for EnKF/GOSI9 system ($4322 \times 3606 \times 75+5$ grid, 24 dynamic + 144 static members, $8\cdot 10^7$ observations, $3.2 \cdot 10^7$ super-observations)}\\
  {\scriptsize$^{(3)}$\,stored in shared memory (one instance per compute node)}\\
  {\scriptsize$^{(4)}$\,when defined as a curvilinear grid}\\
  {\scriptsize$^{(5)}$\,with \verb|STRIDE| = 3}
}

The memory footprint of PREP is defined by the size of the (original, before superobing) \verb|observation| structure array and, in some cases, by the size of curvilinear grids.
The memory usage by curvilinear grids is much reduced by parsing them into K-D trees (default from v1.101.4; the only option since v1.106.0) rather than into binary trees.
Because PREP is not parallelised (mainly due to lack of robust and efficient parallel analogue of \verb|qsort| procedure), in practice its memory footprint is rarely a problem (when running from a master script).

The footprint of CALC is mainly defined by the size of ensemble observations $\mathcal H(\mb E)$.
From version 1.74, it has been substantially reduced for multi-core CPUs by storing only one instance of this array per compute node, with further developments in v1.110.0 and v1.111.11 (see Sec. \ref{sec:flags} on \verb|USE_SHMEM| for details).

For the EnKF, the footprint of UPDATE is mainly defined by the size of the array of horizontal model fields times the number of simultaneously updated fields.
The number of simultaneously updated fields is defined by parameter \verb|FIELDBUFFERSIZE|.
Note that larger values of \verb|FIELDBUFFERSIZE| increase computational effectiveness by reducing the number of reads of and interpolations within $\mb X_5$ arrays.
For the EnOI, the footprint of UPDATE is insensitive to \verb|FIELDBUFFERSIZE| (which should be set to 1), and is defined by the size of the ensemble of horizontal model fields.

For very large models like 1/12-degree NEMO in GOSI9 configuration the ensemble of horizontal fields ($\sim 10$\,GB) can exceed the average memory per core (``slot'').
This requires reducing the number of used slots per NUMA node when running UPDATE.

\subsection{Exit action}

When exiting on an error, EnKF-C by default prints the stack trace, which allows to trace the exit location in the code.
Another option -- to generate a segmentation fault -- can be activated by setting \verb|EXITACTION = SEGFAULT| in the parameter file.
Note that when run on multiple processors, this can result in segmentation faults on more than one processor (but not necessarily on every engaged processor, as some processes can also be forced to exit by \verb|MPI_abort()|).
If the system is set to generate core dumps, they can indeed be used for investigating the final state of the program.

\subsection{Dependencies and compilation issues}

Compiling EnKF-C requires the following external libraries:
\begin{itemize}
\item netcdf (along with hdf5);
\item lapack and blas (or mkl\_rt, which may also require \verb|-lgomp|);
\item MPI (e.g. openmpi), for parallel CALC, UPDATE and ens\_diag.
\end{itemize}
The library settings are specified in file \verb|make.inc| (see the examples in directory \verb|arch|).

EnKF-C also relies on GNU \verb|qsort_r()|, which may be lacking (or have a different signature) in some systems.
In such cases use compile flag \verb|-DINTERNAL_QSORT_R| to activate the internal version of this procedure.

Notes:
\vspace{-3mm}
\begin{enumerate}
\item Using Intel's version of Lapack library -- Intel Math Kernel Library -- can improve performance over Lapack compiled with gfortran.
When running MKL with MPI, it is advisable to use a single thread per process (e.g. \verb|MKL_NUM_THREADS=1|).
\end{enumerate}

\subsection{Random numbers}
\label{sec:seed}

Random numbers are used by EnKF-C for randomisation of model variables (entry \verb|RANDOMISE| in the model parameter file, Sec.~\ref{sec:bias}), for shuffling observations and grid rows (compile flags \verb|OBS_SHUFFLE| and \verb|SHUFFLE_ROWS|), and for randomising the order of inserting nodes into K-D trees.
By default the random number generator is seeded from \verb|/dev/urandom| on the master process, and the seed is broadcast to all processes; therefore these random elements differ from run to run.
To make runs reproducible, set the environment variable \verb|ENKF_SEED| to an integer, e.g.
\begin{Verbatim}
  export ENKF_SEED=12345
\end{Verbatim}
which is then used as the seed (on all processes).

\section{Possible problems / FAQ}

\begin{enumerate}

\item{\bf The code does not compile on OS X platform.}

In some cases compiler can not find \verb|qsort_r()| (or definition of data type \verb|__compar_d_fn_t|), or finds a version with a different signature.
Edit Makefile by adding \verb|-DINTERNAL_QSORT_R| to \verb|PREPCALC_FLAGS|.
This replaces \verb|qsort_r()| by a portable internal implementation in both PREP and CALC, so that no other changes are needed.

\item{\bf CALC quits with ``USE\_SHMEM: MPI ranks on each compute node must be contiguous''.}

The shared-memory code requires block placement of MPI ranks on compute nodes (Sec.~\ref{sec:flags}).
Run \verb|mpirun| with block placement (e.g. \verb|--map-by core| or \verb|--map-by slot|) or recompile CALC without \verb|-DUSE_SHMEM|.

\item{\bf CALC or UPDATE quits with ``directory .enkftmp already exists''.}

This directory is left over by a previous (crashed or interrupted) run; remove it and rerun.

\item{\bf CALC becomes too slow after increasing localisation radius}

This is due to the increased number of local observations.
One way to reduce it is to run superobing on a virtual coarser horizontal grid by increasing parameter \verb|SOBSTRIDE| from the default value of 1 to 2 or more.

\item{\bf CALC takes longer than expected when ``calculating ensemble observations''.}

Make sure that model output is chunked by horizontal layers.

\item{\bf In UPDATE I am getting error \\``\verb|<...> "spread.nc": <...> NC_UNLIMITED size already in use|''.}

UPDATE tries to create a common output file for all model variables.
This may be not straightforward due to possible differences in formats of various model data files involved.
Try compiling UPDATE with flag \verb|-DNCW_SKIPSINGLE|.

\end{enumerate}

\chapter{Setting up a DA system}

This chapter lists the main steps for setting up a geophysical ensemble DA (forecasting) system.
Specifically, it concerns (but not limited to) ocean DA systems.

A DA system is built based on a geophysical model.
Normally, a running model is sufficient for setting up a DA system with EnKF-C.

\section{Initial considerations}

Typical ocean DA systems aim at constraining mesoscale circulation because the available amount of observations and quality of ocean models make it impossible to reliably constrain smaller scales.
Considering that (1) the characteristic error doubling time at mesoscale is of order of a month; (2) unlike variational DA, there is no need for large observation windows in ensemble DA; but (3) smaller cycle lengths incur additional computational overheads, the reasonable DA cycle lengths in ocean DA systems are somewhere between a day and a week.

\section{EnKF systems}
\label{sec:enkf_systems}

\begin{enumerate}

\item{Decide on the ensemble size.}

  In our experience, best performance is achieved with hybrid systems, which also allow smaller ensembles than pure EnKF systems.
  For a pure EnKF system, a typical recommended ensemble size is about 100, although often it can be reduced without substantial deterioration in performance.
  
\item{Create initial ensemble.}

  For the EnKF one needs a ``wide'' initial ensemble such that, ideally, the true system state at the initial time might be considered as a sample from it.
  The detailed composition of the initial EnKF ensemble is not important because it will be ``forgotten'' after the spin-up.
  
  In the context of ocean forecasting the initial ensemble can be created by conducting a long model run (hindcast) and saving model restarts around the time of the year (season) of the initial time.
  The length of this hindcast can be from about 5 to 30 years or more, typically about 10 years.
  The number of model restarts (ensemble members) saved each year is determined by achieving the desired ensemble size.
  Note that if the system is supposed to run with randomly perturbed forcing then it is possible to duplicate some members as they quickly diverge after starting the system.

  A challenging part of the initial hindcast is to obtain possibly larger subsurface variability.
  Typically, the subsurface states of ocean models are substantially different from the true state; therefore larger subsurface variability is important for quicker reduction of initial biases (which can take years).
  Larger subsurface variability can be obtained by longer initial hundcasts; by conducting several hindcasts initialised from different ocean state products; and by perturbing mixing and advection settings.

\item{Decide on the Mean Dynamic Topography (MDT) product.}

  For global systems it normally should be one of the available products that is or can be used in calculating sea level anomaly (SLA) observations.
  For systems with local area models (LAMs) it should be the product used in the system providing the open boundary conditions.

  In some cases the model can be substantially biased about MDT estimates.
  Thes biases project on subsurface state and can cause significant drifts and biases.
  In these cases it may be necessary to use the native MDT calculated as the average sea level during the initial hindcast.

\item{Choose observation time binning, localisation radii and R-factors.}

  Generally, it is preferable to use time-average model outputs for calculating forecast observations.
  The time intervals chosen for particular model variables should be based on the their temporal variability. 
  Ocean circulation at mesoscale evolves very slowly.
  Therefore, the timing for assimilating SLA, temperature and salinity (T and S) observations is not really critical.
  It seems natural to calculate forecast observations using daily average model outputs but it is also possible to use longer intervals or assimilate synchronously using model restarts.

  In contrast, SST exhibits a strong diurnal (daily) cycle.
  Therefore, SST needs to be matched against model outputs on a sub-daily scale.
  Further on, because it is impossible today to reliably model the daytime SST, one needs (1) to restrict assimilation of SST observations to nighttime only and (2) to ensure that no daytime model SST is contributing to the forecast observations.
  Once resticted to nighttime, the SST observations will be matched in DA with SST from the closest in time model dump.
  It is reasonable to use 1- or 2-hour time intervals for SST model outputs.

  Localisation radii and R-factors should be chosen to provide robust inversions in local analyses and sensible increments.
  The same concerns the magnitude of K-factor (Sec.~\ref{sec:kfactor}).

\item{Choose moderation settings.}

  The values of multiplicative inflation and magnitude of RPS (secs.~\ref{sec:capping},\ref{sec:rps}) are set to ensure robust evolution of the DAS, with as small adjustments to the KF (no inflation, no relaxation) as possible.
  To minimise long-term damage from multiplicative inflation it is important to keep it minimal and cap for each element of the state vector by the magnitude of spread reduction (Sec.~\ref{sec:capping}).
  (Capped inflation is a default setting in EnKF-C.)
  See Section~\ref{sec:da-settings-example} for an example of setings from an operational global ocean forecasting system.

\item {DA cycle design.}

  This involves positioning of observation window relative to analysis.
  For EnKF systems it is rather straightforward.
  Assuming that the system is optimal enough, asynchronous DA should work well for an asymmetric cycle (when the analysis is not centred about the observation window).
  In practice, centred observation window may yield somewhat better performance but it is unlikely to be justified by 1.5x increase of the cost of model propagation, which typically constitutes 90--95\,\% of the total cost of the EnKF. 
  Therefore, it is natural to set observation window to the interval from previous to current analysis.

\item{Spin-up of the system}
 
  The system starts from an unconstrained state; therefore, the initial corrections are expected to be rather unbalanced.
  A practical way to spin-up the system is to increase the supposedly optimal R-factors by a large multiple and gradually reduce it in the first several cycles.
  For example, one can use the multiples $2^{7-i}$ for cycle $i, i = 0,\dots,6$ for the first 7 cycles.

\item{Correct the initial state}

  The initial state is represented by the ensemble average of the initial ensemble generated from a long run of the free model.
  This initial state can be substantially biased about SLA, T and S observations.
  In this case it is a good idea to correct it (i.e. subtract the global mean bias from ensemble fields for T and S and add to MDT for SSH); otherwhile it can take a long time (of order of a year) to get it reduced by DA.
  
\end{enumerate}

\section{EnOI}

\begin{enumerate}
  
\item{Create static ensemble.}

  Unlike the inital ensemble for an EnKF system, which gets ``forgotten'' over time, the static ensemble in an EnOI system does not evolve and therefore continuosly affects the performance of the system.
  Because of its static nature, it cannot represent the ``errors of the day'' and should aim at representing large scale variability.

  The EnOI ensemble is created based on a long model run.
  In contrast to the EnKF, the main outputs should be not model restarts but time average fields for the updated variables.
  The ensemble should not contain long term trends including seasonal trends.
  This can be accomplished by generating ensemble members as the difference between time averaged model fields over two time windows of different width centered about each other.
  The particular lengths of these windows can be optimised depending on the model resolution and domain.
  Typically, for ocean forecasting, the inner window can be from 1-2 days to a week, and the outer window from 1-2 weeks to 1-3 months.
  In practice, it is convenient to save 1-day model average fields during the hindcast and then generate the ensemble based on this output.

  The size of the static ensemble can be typically about 200.
  Larger ensembles start affecting the cost of matrix inversions (that scale as the size of the matrix cubed) while bringing diminishing benefits for the system performance.

  Note that for ocean EnKF systems it is desirable to exclude the atmospheric component of SSH (``inverse barometer'') from the saved fields; otherwise the ensemble SSH anomalies would contain this extra variability, which negatively affects the balances between model fields in the ensemble.
  (Although the suggested way of calculating ensemble anomalies as the difference of time average in two centred windows can largely eliminate the atmospheric variability in SSH.)
  
\item {DA cycle design.}

  The EnOI is essentially an interpolation technique and, as such, produces increments for the centre of the observation window.
  Further, because EnOI systems propagate only one instance of the model, the cost of model propagation is not as critical as for the EnKF.
  It is therefore natural to centre the observation window about the analysis time.

  With FGAT approach this means that at each DA cycle the model needs to be propagated for the half of DA cycle to the start of the next observation window and then further for the full cycle to the end of the observaiton window, i.e. for 1.5 DA cycle in total.
  The model needs to save restart at the time of the next analysis, i.e. after propagating for 1.0 DA cycle.

\item {DA settings.}

  Generally, EnOI systems reach best performance with somewhat stronger fit to observations than EnKF systems (with smaller R-factors and larger localisation radii than in EnKF systems with ensembles of similar spread).
  Note that the ensemble spread depends on the ensemble generation technique and  settings, which involves additional tuning by means of a multiple to R-factors.
  
\end{enumerate}

\section{Hybrid EnKF/EnOI}

Today hybrid EnKF/EnOI systems represent the state-of-the art in the ensemble based ocean forecasting.
For example, OceanMAPS v4.0 and v4.1 are global (75\,S to 75\,N) eddy resolving (0.1$^\circ$) ocean forecasting systems run by Australian Bureau of Meteorology.
They are hybrid systems with 48 dynamic and 144 static ensemble members.
These systems considerably outperform otherwise similar EnKF systems with 96 ensemble members at half of their computational cost.

Setting up a hybrid system involves the following steps.

\begin{enumerate}

\item {Generate static and dynamic ensemble.}

  Follow steps described in the previous sections for generating EnKF and EnOI ensembles.

\item{Decide on the scaling coefficient \verb|GAMMA| of the static ensemble.}

  In practice OceanMAPS does not show much sensitivity to \verb|GAMMA|.

\item{DA cycle design.}

  It seems natural to follow the design described for the EnKF.
  In situations when computing expenses are not an issue it is also possible to use observation window symmetric about analysis.

\end{enumerate}

\subsection{Settings in OceanMAPS v4.1}
\label{sec:da-settings-example}

\begin{tabular}{ll}
  \bf Ensemble size: & 48 dynamic members, 144 static members\\
  \bf Cycle length: & 1\,d\\
  \bf Observation window: & $[-1, 0]$\,d about analysis\\
  \bf Observation binning: & SLA 1\,d, SST 2\,h, T 1\,d, S 1\,d\\
  \bf R-factors: & SLA 4.5, SST 48, T 18, S 18\\
  \bf Loc. support radii (km): & SLA 175, SST 150, T 450, S 450\\
  \bf ALPHA: & 0.70\\
  \bf GAMMA: & 0.25\\
  \bf Inflation: & 1.02 1\\
  \bf K-factor: & 2\\
\end{tabular}

\chapter*{Acknowledgements}
\addcontentsline{toc}{chapter}{Acknowledgments}

EnKF-C has been developed during author's work with Bureau of Meteorology on Bluelink project.
The author has used his knowledge of TOPAZ \citep{sak12b} and BODAS \citep{oke08b} systems and borrowed from them a number of design solutions and features.
Paul Sandery was the first user of this code (apart from the author), and his enthusiastic support is cheerfully acknowledged.

\clearpage

\nocite{eve94a}
\nocite{eve03a}
\nocite{hun04a}
\nocite{hun07a}
\nocite{sak08a}
\nocite{sak10a}
\nocite{sak11a}

\bibliographystyle{ametsoc}
\bibliography{enkf}

\begin{thebibliography}{22}
\expandafter\ifx\csname natexlab\endcsname\relax\def\natexlab#1{#1}\fi
\expandafter\ifx\csname url\endcsname\relax
  \def\url#1{{\tt #1}}\fi
\expandafter\ifx\csname urlprefix\endcsname\relax\def\urlprefix{URL }\fi
\expandafter\ifx\csname doiprefix\endcsname\relax\def\doiprefix{doi:}\fi

\bibitem[{Anderson(2003)}]{and03a}
Anderson, J.~L., 2003: A local least squares framework for ensemble filtering.
  {\it Mon. Wea. Rev.\/}, {\bf 131}, 634--642.

\bibitem[{Andrews(1968)}]{and68a}
Andrews, A., 1968: A square root formulation of the {K}alman covariance
  equations. {\it AIAA J.\/}, {\bf 6}, 1165--1168.

\bibitem[{Bishop et~al.(2001)Bishop, Etherton, and Majumdar}]{bis01a}
Bishop, C.~H., B.~Etherton, and S.~J. Majumdar, 2001: Adaptive sampling with
  the ensemble transform {K}alman filter. part {I}: theoretical aspects. {\it
  Mon. Wea. Rev.\/}, {\bf 129}, 420--436.

\bibitem[{Bocquet(2016)}]{boc16a}
Bocquet, M., 2016: Localization and the iterative ensemble {K}alman smoother.
  {\it Q. J. R. Meteorol. Soc.\/}, {\bf 142}, 1075--1089.

\bibitem[{Evensen(1994)}]{eve94a}
Evensen, G., 1994: Sequential data assimilation with a nonlinear
  quasi-geostrophic model using {M}onte-{C}arlo methods to forecast error
  statistics. {\it J. Geophys. Res.\/}, {\bf 99}, 10143--10162.

\bibitem[{Evensen(2003)}]{eve03a}
--- 2003: The {E}nsemble {K}alman {F}ilter: theoretical formulation and
  practical implementation. {\it Ocean Dynam.\/}, {\bf 53}, 343--367.

\bibitem[{Evensen(2004)}]{eve04a}
--- 2004: Sampling strategies and square root analysis schemes for the {EnKF}.
  {\it Ocean Dynam.\/}, {\bf 54}, 539--560.

\bibitem[{Evensen and van Leeuwen(2000)}]{eve00a}
Evensen, G. and P.~J. van Leeuwen, 2000: An ensemble {K}alman smoother for
  nonlinear dynamics. {\it Mon. Wea. Rev.\/}, {\bf 128}, 1852--1867.

\bibitem[{Hunt et~al.(2004)Hunt, Kalnay, Kostelich, Ott, Patil, Sauer,
  Szunyogh, Yorke, and Zimin}]{hun04a}
Hunt, B.~R., E.~Kalnay, E.~J. Kostelich, E.~Ott, D.~J. Patil, T.~Sauer,
  I.~Szunyogh, J.~A. Yorke, and A.~V. Zimin, 2004: Four-dimensional ensemble
  {K}alman filtering. {\it Tellus\/}, {\bf 56A}, 273--277.

\bibitem[{Hunt et~al.(2007)Hunt, Kostelich, and Szunyogh}]{hun07a}
Hunt, B.~R., E.~J. Kostelich, and I.~Szunyogh, 2007: Efficient data
  assimilation for spatiotemporal chaos: A local ensemble transform {K}alman
  filter. {\it Physica D\/}, {\bf 230}, 112--126.

\bibitem[{Kalman(1960)}]{kal60}
Kalman, R.~E., 1960: A new approach to linear filtering and prediction
  problems. {\it J. Basic. Eng.\/}, {\bf 82}, 35--45.

\bibitem[{Oke et~al.(2008)Oke, Brassington, Griffin, and Schiller}]{oke08b}
Oke, P.~R., G.~B. Brassington, D.~A. Griffin, and A.~Schiller, 2008: The
  {B}luelink ocean data assimilation system ({BODAS}). {\it Ocean Model.\/},
  {\bf 21}, 46--70.

\bibitem[{Ott et~al.(2003, rev. 2005)Ott, Hunt, Szunyogh, Zimin, Kostelich,
  Corazza, Kalnay, Patil, and Yorke}]{ott03a}
Ott, E., B.~R. Hunt, I.~Szunyogh, A.~V. Zimin, E.~J. Kostelich, M.~Corazza,
  E.~Kalnay, D.~J. Patil, and J.~A. Yorke, 2003, rev. 2005: A local ensemble
  {K}alman filter for atmospheric data assimilation. {\it
  http://arxiv.org/abs/physics/0203058\/}.

\bibitem[{Sakov and Bertino(2011)}]{sak11a}
Sakov, P. and L.~Bertino, 2011: Relation between two common localisation
  methods for the {EnKF}. {\it Comput. Geosci.\/}, {\bf 15}, 225--237.

\bibitem[{Sakov et~al.(2012)Sakov, Counillon, Bertino, Lis{\ae}ter, Oke, and
  Korablev}]{sak12b}
Sakov, P., F.~Counillon, L.~Bertino, K.~A. Lis{\ae}ter, P.~R. Oke, and
  A.~Korablev, 2012: {TOPAZ4}: an ocean-sea ice data assimilation system for
  the {N}orth {A}tlantic and {A}rctic. {\it Ocean Science\/}, {\bf 8},
  633--656.

\bibitem[{Sakov et~al.(2010)Sakov, Evensen, and Bertino}]{sak10a}
Sakov, P., G.~Evensen, and L.~Bertino, 2010: Asynchronous data assimilation
  with the {EnKF}. {\it Tellus\/}, {\bf 62A}, 24--29.

\bibitem[{Sakov and Oke(2008{\natexlab{a}})}]{sak08a}
Sakov, P. and P.~R. Oke, 2008{\natexlab{a}}: A deterministic formulation of the
  ensemble {K}alman filter: an alternative to ensemble square root filters.
  {\it Tellus\/}, {\bf 60A}, 361--371.

\bibitem[{Sakov and Oke(2008{\natexlab{b}})}]{sak08b}
--- 2008{\natexlab{b}}: Implications of the form of the ensemble
  transformations in the ensemble square root filters. {\it Mon. Wea. Rev.\/},
  {\bf 136}, 1042--1053.

\bibitem[{Sakov and Sandery(2017)}]{sak17a}
Sakov, P. and P.~Sandery, 2017: An adaptive quality control procedure for data
  assimilation. {\it Tellus A\/}, {\bf 69}, 1318031.

\bibitem[{Verlaan and Heemink(1997)}]{ver97a}
Verlaan, M. and A.~W. Heemink, 1997: Tidal flow forecasting using reduced rank
  square root filters. {\it Stoch. Hydrol. Hydraul.\/}, {\bf 11}, 349--368.

\bibitem[{Yang et~al.(2009)Yang, Kalnay, Hunt, and Bowler}]{yan09a}
Yang, S.-C., E.~Kalnay, B.~Hunt, and E.~N. Bowler, 2009: Weight interpolation
  for efficient data assimilation with the {Local Ensemble Transform Kalman
  Filter}. {\it Q. J. R. Meteorol. Soc.\/}, {\bf 135}, 251--262.

\bibitem[{Zhang et~al.(2004)Zhang, Snyder, and Sun}]{zha04a}
Zhang, F., C.~Snyder, and J.~Sun, 2004: Impacts of initial estimate and
  observation availability on convective-scale data assimilation with an
  ensemble {K}alman filter. {\it Mon. Wea. Rev.\/}, {\bf 132}, 1238--1253.

\end{thebibliography}
\addcontentsline{toc}{chapter}{References}

\clearpage

\chapter*{Abbreviations}
\addcontentsline{toc}{chapter}{Abbreviations}

Used in data assimilation:\\[2mm]
\begin{tabular}{lll}
  CL &-& covariance localisation \\
  DEnKF &-& deterministic EnKF \\
  DA &-& data assimilation \\
  DAS &-& data assimilation system \\
  DAW &-& data assimilation window \\
  DFS &-& degrees of freedom of signal \\
  EKF &-& extended Kalman filter \\
  EnKF &-& ensemble Kalman filter \\
  EnOI &-& ensemble optimal interpolation \\
  ETKF &-& ensemble transform Kalman filter \\
  ETM &-& ensemble transform matrix \\
  FGAT &-& first guess at appropriate time \\
  KF &-& Kalman filter \\
  KS &-& Kalman smoother \\
  LA &-& local analysis \\
  QC &-& quality control \\
  RPS &-& relaxation to prior spread \\
  SDAS &-& state of data assimilation system \\
  SLA &-& sea level anomalies \\
  SRF &-& spread reduction factor \\
  SST &-& sea surface temperature \\
  SVD &-& singular value decomposition \\
\end{tabular}

Used in ocean modelling:\\[2mm]
\begin{tabular}{lll}
  LAM &-& local area model \\
  MDT &-& mean dynamic topography \\
  MLD &-& mixed layer depth \\
  SIC &-& sea-ice concentration \\
  SIT &-& sea-ice thickness \\
  SLA &-& sea level anomaly \\
  SSH &-& sea surface height \\
  SST &-& sea surface temperature\\
\end{tabular}

\clearpage

\chapter*{Symbols}
\addcontentsline{toc}{chapter}{Symbols}

\section*{General symbols}
\begin{tabular}{lll}
  $\mb x$ (small, bold) &-& a vector \\
  $\mb 1$ &-& a vector with all elements equal to 1 \\
  $\mb 0$ &-& a vector with all elements equal to 0 \\
  $\mb A$ (capital, bold) &-& a matrix \\
  $\mb I$ &-& an identity matrix \\
  $\mb U$ &-& a unitary matrix, $\mb U \mb U\T = \mb I$ \\
  $\mb A\T$ &-& transposed matrix $\mb A$ \\
  $\mb A^{1/2}$ &-& the unique positive definite square root of a positive definite matrix $\mb A$ \\
  $\mb A(m_1 : m_2, n_1 : n_2)$ &-& the block of $\mb A$ composed of rows from $m_1$ to $m_2$ and columns from $n_1$ to $n_2$\\
  $\mathrm{tr}(\mb A)$ &-& trace of $\mb A$ \\
  $\mathcal H \circ \mathcal M(\mb x)$ &-& $\mathcal H \left[ \mathcal M (\mb x) \right]$ \\
  $\mb A \circ \mb B$ &-& by-element, or Hadamard, or Schur product of matrices\\
  $\|\mb x\|^2_{\mb B }$ &-& $ \mb x\T \mb B \mb x$
  \end{tabular}

\section*{DA related symbols}
\begin{tabular}{lll}
  $m$ &-& ensemble size \\
  $n$ &-& state size \\
  $p$ &-& number of observations \\
  $\mb A$ &-& ensemble anomalies, $\mb A = \mb E - \mb x \mb 1\T$ \\
  $\mb E$ &-& ensemble \\
  $\mb G$ &-& an intermediate matrix in the EnKF analysis, $\mb G \equiv (\mb I + \mb S\T \mb S)^{-1} \mb S\T = \mb S\T ( \mb I + \mb S \mb S\T)^{-1}$ \\
  $\mathcal H$ &-& nonlinear observation operator; in linear case -- affine observation operator \\
  $\mb H$ &-& linearised observation operator, $\mb H = \nabla \mathcal H(\mb x)$ \\
  $J$ &-& cost function \\
  $\mathcal M$ &-& nonlinear model operator; in linear case -- affine model operator \\
  $\mb M$ &-& linearised model operator, $\mb M = \nabla \mathcal M(\mb x)$ \\
  $\mb P$ &-& state error covariance estimate; also used as abbreviation for $\mb A \mb A\T / (m - 1)$ \\
  $\mb Q$ &-& model error covariance \\
  $\mb R$ &-& observation error covariance \\
  $\mb S$ &-& normalised ensemble observation anomalies, $\mb S = \mb R^{-1/2} \mb {HA} / \sqrt{m - 1}$ \\
  $\mb T_L$ &-& left-multiplied ensemble transform matrix, $\mb A^a = \mb T_L \mb A^f$ \\
  $\mb T_R$ &-& right-multiplied ensemble transform matrix, $\mb A^a = \mb A^f \mb T_R$ \\
  $\mb U^p$ &-& a unitary mean-preserving matrix, $\mb U^p (\mb U^p) \T = \mb I$, $\mb U^p \mb 1 = \mb 1$ \\
  $\mb X_5$ &-& historic symbol for the full ensemble transform matrix, $\mb E^a = \mb E^f \mb X_5$ \\ 
  $\mb s$ &-& normalised innovation, $\mb s = \mb R^{-1/2} \left[ \mb y - \mathcal H (\mb x^f) \right] / \sqrt{m - 1}$ \\
  $\mb x$ &-& state estimate \\
  $\mb y$ &-& observation vector \\
  $\mb w$ &-& vector of linear coefficients for updating the mean, $\mb x^a = \mb x^f + \mb A^f \mb w$ \\
  $(\cdot)^f$ &-& forecast expression \\
  $(\cdot)^a$ &-& analysis expression \\
  $(\cdot)_i$ &-& either expression at cycle $i$ or $i$th element of a vector \\
  $\ac{i}{(\cdot)}$ &-& local expression for state element $i$ \\
  $\ac{\{o\}}{(\cdot)}$ &-& local expression for observation $o$
\end{tabular}

\end{document}